# Quantum Defects in 2D Transition Metal Dichalcogenides for Terahertz Technologies

Jingda Zhang[1] and Su Ying Quek[1,2,3,4,5*]

[1]Department of Physics, National University of Singapore, Singapore 117551

[2]Centre for Advanced 2D Materials, National University of Singapore, Singapore 117546

[3]Department of Materials Science and Engineering, National University of Singapore, Singapore 117575

[4]Integrative Sciences and Engineering Programme, NUS Graduate School, National University of Singapore, Singapore 119077

[5]NUS College, National University of Singapore, Singapore 138593

*Corresponding author: S.Y.Q. (email: phyqsy@nus.edu.sg)

## ABSTRACT

Substitutional transition metal (TM) point defects have recently been controllably introduced in two-dimensional (2D) transition metal dichalcogenides. We identify quantum defect candidates through a first-principles materials discovery approach with 25 TM elements substituting Mo and W in 2D $MoS_2$ and $WSe_2$, respectively. We elucidate trends in the charge transition levels for these 50 systems and report the existence of defects with spin-triplet ground states and a zero-field splitting (ZFS) in the terahertz (THz) regime, in contrast to typical gigahertz values. These defects can couple to resonant near-infrared radiation, providing a route to applications as high-fidelity qubits controlled by spin-dependent optical transitions. The THz ZFS implies that these high-fidelity operations can take place at higher temperatures compared to the case for GHz qubits. Our results also point toward the possibility of realising a single-photon THz emitter. This work broadens the scope of quantum defects, highlighting the opportunities for next generation THz quantum technologies – an area of growing interest given the rapid advancement in the development of THz sources and detectors.

## INTRODUCTION

Point defects introduce atomic-like electronic states within the band gap of host materials. Defects with optically addressable spins can function as qubits, the building block for quantum information technologies.[1–3] There are two major approaches[3] by which light can be used to control or read out the spin state, i.e. the spin sublevel or magnetic quantum number. First, non-radiative processes that are dependent on the spin state can lead to selective population of a spin sublevel, through intersystem crossing (ISC) (coupling between states with different total spin angular momentum, e.g. singlet ($S = 0$) and triplet ($S = 1$)). This technique enables control and read-out of the spin state, through continuous cycles of optical pumping. A well-known application is the room temperature control of the nitrogen-vacancy charge centre ($NV^-$) in diamond.[4] The second approach for the optical control of a spin qubit is based on well-established techniques in atomic physics – resonant excitation of the qubit using spin-conserving optical transitions that have different frequencies or selection rules for different spin sublevels.[5] Initialisation of the qubit takes place through projective measurements,[5] with the option of manipulating the state through Rabi oscillations between the spin sublevels. This approach has the advantage of enabling high fidelity operation of the spin qubit, as demonstrated in the group-IV colour centres and also in the $NV^-$ centre in diamond.[5–7] The price that is paid in exchange for higher fidelity operations is the lower temperature at which such qubits operate; in the case of $NV^-$, a temperature of less than 10 K was used,[5] and these low temperatures limit the scaling of this technology.

A clear route to increasing the operation temperature for the second approach described above, is to increase the difference in resonant frequencies for the different optical transitions, and to increase the energy difference between the spin sublevels. A larger energy splitting also reduces phonon-mediated transitions between spin sublevels,[6] translating into prolonged spin relaxation times. Spin coherence times are limited by the spin relaxation times,[3] and furthermore, if dynamical decoupling techniques are applied,[8,9] the spin coherence time is also phonon-limited.[10] Thus, reducing the phonon-mediated transitions can allow for more robust quantum memory technologies at higher temperatures.[11] The quantity of interest is the zero-field splitting (ZFS), which represents the energy difference between spin sublevels in the absence of a magnetic field. For commonly studied spin defects such as the $NV^-$ centre in diamond and its analogues,[12,13] the ZFS is dominated by spin-spin dipole-dipole interactions and is typically in the GHz range.[14,15]

ZFS in the THz range has been reported in transition metal (TM) complexes and molecules in the literature,[16,17] where such large ZFS arises from strong spin-orbit coupling (SOC). The large ZFS has been of interest for applications of these molecules as single molecule magnets for quantum information processing at room temperature. Are there defects hosted in solid-state systems that can have a spin-triplet ground state and a ZFS in the THz range, and if so, are they amenable to resonant optical control as solid-state spin qubits? Another intriguing question, tangential to the discussion above, is whether or not it is possible for the THz ZFS in these defects to be used to generate single THz photons, so as to realise a THz quantum emitter. A necessary condition for the THz quantum emitter concept is the ability to achieve population inversion between the spin sublevels.

We will investigate these questions in substitutional TM point defects in two-dimensional (2D) transition metal dichalcogenides (TMDs). 2D solid-state materials offer significant advantages for single photon emitters and high-performance optically-addressable defect systems due to their ease of integration with optoelectronic and nanophotonic components.[18–20] Quantum emitters in 2D materials also benefit from a low level of total internal reflection, leading to higher light extraction efficiencies, and the possibility of enhancing the emission strength by integration with cavities[21,22] and photonic waveguides or plasmonic structures.[23] Among the 2D materials, 2D TMDs in particular are promising host platforms[20] because of their semiconducting nature which allows for gate-tunability[24] to access multiple stable charge states, and the existence of scalable and highly controlled strategies for uniformly incorporating point defects involving substitutional TM atoms in 2D TMDs.[25–27]

Based on first-principles density functional theory (DFT) calculations, we identify potential quantum defect candidates using a high-throughput materials discovery approach in which 25 different TM elements are incorporated into $MoS_2$ and $WSe_2$, replacing single Mo or W atoms, respectively. Clear trends in the charge transition levels (CTLs) across the periodic table are established, and the stability of spin and charge states for these 50 different systems is elucidated. We show that these point defects have remarkably large ZFS, typically in the sub-THz to THz range. Importantly, we identify point defects that have stable spin-triplet states, with sub-THz to THz ZFS, but with minimal spin mixing due to a significant singlet-triplet energy difference. These defects can furthermore be controlled by resonant near-infrared wavelengths. We therefore identify potential candidates for spin qubits in 2D TMDs, with a ZFS in the THz range, which can overcome the limitation of cryogenic temperatures for high-fidelity resonant excitation control of the qubit. We also find one system, the neutral Fe substitutional defect in $MoS_2$, which can potentially be a THz quantum emitter. This work is timely given the recent rapid advancements in the development of THz sources and detectors,[28,29] and identifies potential quantum defects in 2D materials, for the new and emerging field[11,30] of THz quantum technologies.

# RESULTS

**Screening of transition metal (TM) substitutional defects in monolayer TMDs**

The pristine monolayer (ML) 2H-TMDs, $MoS_2$ and $WSe_2$, exhibit $D_{3h}$ point group symmetry with each Mo or W atom having a trigonal prismatic coordination with adjacent chalcogen atoms. To create substitutional point defects, we replace one Mo or W atom in a 4×4×1 supercell with a 3d, 4d or 5d TM element from group IVB to IIB across the periodic table (excluding the radioactive technetium) (Figure 1a), amounting to 25 different TM dopants considered in each of $MoS_2$ and $WSe_2$. Details of our first principles approaches are found in the Methods section of the Supporting Information (SI), and a summary is shown in SI, Table S1.

We first wish to investigate the stable charge states of these systems, as a starting point to search for defects with spin-triplet ground states – the simplest spin configuration that exhibits ZFS. The introduction of the TM substitutional defect with different charge states can sometimes disrupt the original trigonal prismatic coordination due to changes in the positions of the TM dopant and adjacent chalcogen atoms. Such Jahn-Teller (JT) or pseudo Jahn-Teller (PJT) distortion has also been reported in the literature.[31–34] A systematic study with geometry optimisation starting from all possible point groups for the 50 different systems with at least three charge states (neutral, −1 and +1), for up to two spin states, requires significant computational resource. In particular the point groups to be considered are $D_{3h}$, $C_{3v}$, $C_{2v}$, $C_s$, $C_h$, and $C_1$ with two distinct possibilities each for $C_{2v}$ and $C_s$ (see SI, Note S1). After extensive investigations, we propose a geometry optimisation procedure starting from the lowest symmetry starting point, $C_1$, in our initial screening process for each of the charge states in the 50 different systems. Using this approach, the point group symmetries of the TM defects are determined. Among the 50 TM dopants in three charge states, we identify 61 systems with $D_{3h}$ symmetry, 30 with $C_{2v}$ symmetry, 30 with $C_s$ symmetry, 22 with $C_1$ symmetry, 5 with $C_h$ symmetry, and 2 with $C_{3v}$ symmetry (SI, Table S2). Among these, the geometries of defects with spin-triplet ground states follow the group symmetries of $D_{3h}$ (no displacement), $C_{2v}$ (in-plane displacement) and $C_1$ (out-of-plane displacement), the symmetry operations of which are shown in Figure 1b.

Once potential qubits are identified, the potential energy landscape is explored more systematically, with geometry optimisation starting from structures with different point group symmetries. For two representative spin qubit candidates, $Mn^{1-}$ and $Fe^{0}$ in ML $MoS_2$, we have found that the lowest energy configurations obtained by explicit consideration of different symmetries are the same as those obtained by relaxation from $C_1$ (SI, Note S1). Thus, the screening approach with geometry optimisation starting from $C_1$ represents a compromise between accuracy and prudence in our use of computational resource for the screening process. Furthermore, while point group symmetries have significant implications for selection rules in optical transitions, the energy gain from the geometric distortions is small compared to the energy scales involved in ionization energies of the defects, which are the important quantities in our initial screening step.

For the screening process, we use the PBE exchange-correlation functional,[35] and an on-site Coulomb interaction U of 3 eV assigned to all the 3d dopants.[36] Further studies on the energy landscape are performed using PBE+U with SOC, as well as using the HSE06 functional[37] that is typically used for obtaining the total energies and eigenvalues for defects.[38,39] While a recent study[40] points out some limitations of PBE+U in 2D materials, we emphasize that all key results in this work are based on HSE06 calculations.

Our studies on the stable charge states in these TM substitutional defects reveal meaningful trends which we now describe. The stability of a charge state depends on the value of the Fermi level relative to the thermodynamic CTLs. Starting from the first group to the right of Mo and W (vertical lines in Figure 2), the (0/–1) and (+1/0) CTLs generally shift down as the group corresponding to the TM dopant shifts right across the periodic table to Group IIB, and this trend continues as we loop back to Group IVB, followed by Group VB (Figure 2). These trends make sense because dopants with fewer valence electrons than elements in group VIB (Cr/Mo/W) have a tendency to accept electrons in $MoS_2$ and $WSe_2$, while those with more valence electrons tend to lose electrons. Nevertheless, it is evident that, for a considerable number of elements, the neutral, –1 and +1 charge states can all be stabilized with an appropriate Fermi level in the band gap.

**Spin-triplet ground states**

To identify defects with spin-triplet ground states, we compute the energy differences $\Delta E$ between the spin-singlet and spin-triplet states of the defect systems with an even number of electrons. We compute $\Delta E$ and identify 8 defect systems that are verified to have a stable spin-triplet state using both PBE+U and HSE06 functionals (Table 1): $Mn^{1-}$, $Fe^{0}$, and $Cu^{1-}$ in both $MoS_2$ and $WSe_2$, as well as $Ti^{0}$ and $Cd^{0}$ in $WSe_2$. $Mn^{1-}$ and $Fe^{0}$ introduce two extra unpaired electrons into the system, while $Ti^0$ lacks two electrons and thus has two unpaired holes. $Cu^{1-}$ has two paired and two unpaired electrons. $Cd^0$ has filled orbitals in its electronic configurations with the two highest occupied unpaired electrons.

Table 1 shows the point group symmetries and singlet-triplet energy differences of the 8 defect candidates obtained using HSE06. Henceforth, we shall use the HSE06 functional for all calculations.

**Zero-field splitting in the terahertz range**

We now quantify the splitting between spin sublevels in the spin-triplet states of these 8 defect systems. In the absence of external magnetic fields, the ZFS causes the three-fold degeneracy of the spin-triplet state to be lifted. The spin sublevel eigenstates in the presence of ZFS are the $m_s = 0$ state, labelled by $|0\rangle$, as well as linear combinations, $|+\rangle$ and $|-\rangle$, of the $m_s = \pm 1$ states. $|+\rangle$, $|-\rangle$, and $|0\rangle$ take the form of $\frac{1}{\sqrt{2}}(|\uparrow\uparrow\rangle + |\downarrow\downarrow\rangle)$, $\frac{1}{\sqrt{2}}(|\uparrow\uparrow\rangle - |\downarrow\downarrow\rangle)$, and $\frac{1}{\sqrt{2}}(|\uparrow\downarrow\rangle + |\downarrow\uparrow\rangle)$, respectively, where $|\uparrow\rangle$ and $|\downarrow\rangle$ are along the spin quantisation axis. Their corresponding eigenenergies are $\frac{1}{3}D + E$, $\frac{1}{3}D - E$, and $-\frac{2}{3}D$, respectively, where $D$, the axial component, represents the energy difference between the $|0\rangle$ sublevel and the average energy of the $|+\rangle$ and $|-\rangle$ sublevels, while the rhombic component $E$ describes the energy difference between the $|+\rangle$ and $|-\rangle$ sublevels.[41] We refer the reader to the Methods section where we derive expressions for the ZFS parameters, $D$ and $E$, in terms of energy differences for different spin quantisation axes. In addition to the contribution to the ZFS from spin-orbit interactions, the spin-spin contribution to the ZFS is also computed. The final spin quantisation axes are obtained by considering both spin-orbit and spin-spin contributions; in most of the systems here, the magnitude of the ZFS from spin-orbit interactions is significantly larger than that from spin-spin interactions. As illustrated in Figure 3a, if the magnetization prefers to align along the spin quantisation axis (i.e., $D^{SO} < 0$, $m_s = \pm 1$ states are more stable), it can be referred to as easy-axis magnetization; in contrast, easy-plane anisotropy describes the case when the magnetization aligns within a plane perpendicular to the spin quantisation axis ($D^{SO} > 0$).

Similar to other systems such as defects in diamond and silicon carbide (SiC),[14] we obtain spin-spin contributions to the ZFS in the GHz range. The large SOC in these systems results in a large spin-orbit contribution to the ZFS which we find to be typically in the sub-THz to THz range (Figure 3b). The axial and rhombic ZFS components, including contributions from both spin-spin and spin-orbit interactions, are provided in the SI for all systems with spin triplet ground states (Table S5). The ZFS values are particularly large for the TM substitutional defects in $WSe_2$, because the defect orbitals hybridize with the neighbouring Mo/W (and chalcogen) atoms (see Figures 4c, 5c and 6c below), and SOC is more significant in $WSe_2$ compared to $MoS_2$ due to the presence of heavier elements. These ZFS values exceed those commonly reported in the literature by more than an order of magnitude – for instance, the $NV^-$ centre in diamond and spin defects in SiC and hexagonal boron nitride (h-BN) typically have ZFS values of less than 10 GHz.[4,42,43] Compared to previous computational studies, the ZFS values of our TM substitutional defects are significantly larger than that of the Ti dopant substituting the divacancy site of ML h-BN with a ZFS of 19.4 GHz,[44] the $Cr_N^-$ defect in ML h-BN with a ZFS of 25.7 GHz,[45] and the defect, $Mo_S$, in ML $WS_2$, which has a ZFS value of 21.7 GHz.[46]

We also note that the ZFS can be controlled by strain engineering. For example, we present the variation of the ZFS parameter, $D^{SO}$, as a function of biaxial in-plane strain in $Fe^0$-$MoS_2$ (SI, Figure S2). A biaxial in-plane strain of within ±2% changes the $D^{SO}$ value in $Fe^0$-$MoS_2$ by ~15%. The predicted strain-dependent ZFS values offer possibilities of a tunable THz quantum information device.

**THz spin qubits**

Resonant optical excitations allow for higher-fidelity operations compared to the use of ISC approaches to control the qubits.[5] The operation of GHz qubits using resonant optical excitations typically requires a cryogenic condition; the very low temperature allows the resonant transitions to be spectrally resolved[5] and suppresses phonon-mediated spin transitions. The ZFS being larger by several orders of magnitude greatly weakens phonon-mediated sublevel transitions and enables a higher temperature for qubit initialization.[6,11] Specifically, phonon-mediated spin transition rates are proportional to the occupation of the phonons with frequencies resonant with the ZFS; increasing the ZFS from 1 GHz to 1 THz decreases the phonon occupation by three orders of magnitude (for a given temperature). Resonant spin-selective optical excitation can be used for high-fidelity spin readout from the frequency or intensity dependence of the emission peaks, as shown previously for GHz qubits.[5,12] The spectral resolution for distinguishing the resonant excitations in different spin channels will also be substantially enhanced with a THz ZFS, so that operation at higher temperatures, where phonon-limited line-widths are present, can become feasible.

Interest in THz quantum technologies and THz qubits is emerging.[11] In a recent review paper,[11] it was recognised that research in THz technologies is gaining momentum in many applications (see discussion below under "THz technologies") but is trailing behind in the quantum domain. In particular, it was emphasised in Ref. 11 that quantum technologies that operate in the THz range can increase the operating temperature of these qubits, and simplify the cooling equipment required for operations.

A THz ZFS also potentially increases the spin relaxation and spin coherence times[3,6] for a given temperature, especially with the application of dynamical decoupling techniques.[8,9] Therefore, a THz spin qubit can be operated with longer spin relaxation and coherence times compared to a GHz spin qubit, for the same temperature, and this enables more complex qubit operations to be performed.

**Large singlet-triplet energy gaps and well-defined triplet states**

Another consideration in our identification of THz spin qubits was the stability of the spin-triplet ground state. Ideally, our spin qubits should have spin-triplet ground states with significant singlet-triplet energy gaps, and from Table 1, we identify the four most stable spin-triplet systems to be the $Mn^{1-}$ and $Fe^0$ defects.

Focusing on the $Mn^{1-}$ and $Fe^0$ defects, we perform HSE06 calculations including SOC self-consistently to evaluate whether the triplet state retains a spin of $S = 1$ in the presence of SOC. We find that the electron spin in these systems remains fully polarized and localized, with a total magnetic moment of 2.0 $\mu_B$ in each system, consistent with a triplet state. In contrast, the spin-triplet systems with a small singlet-triplet energy gap have a total magnetic moment that is significantly smaller than 2.0 $\mu_B$ when SOC is included in the calculations – for example, Ti and Cd in $WSe_2$ are predicted to have total magnetic moments of 1.4 and 1.8 $\mu_B$, respectively. We therefore turn our attention to the $Mn^{1-}$ and $Fe^0$ defects.

**Selected spin qubit candidates**

Single-particle band structures are calculated at the HSE06 level for the quantum defect candidates, $Mn^{1-}$ and $Fe^0$ in $MoS_2$ and $WSe_2$ (see SI, Figure S3 and S4 for their atomic structures and SI, Figure S5 and S6 for their band structures in both 4×4 and 6×6 supercells). As $Fe^0$-$WSe_2$ has defect states that are quite close to the pristine valence bands for $WSe_2$, we focus on the remaining three systems.

**$Fe^0$-$MoS_2$**. $Fe^0$-$MoS_2$ exhibits $D_{3h}$ symmetry (see local geometry in Figure 4a), and the single-particle defect levels and corresponding wavefunctions are shown in Figure 4b and 4c. The irreducible representations (IRs) of the many-body spin sublevels are shown in Figure 4d. The Fe $d$-orbitals hybridize with adjacent Mo $d$-orbitals and S $p$-orbitals. Due to the large exchange splitting, the only in-gap defect states are spin-up states, which have bonding characteristics. The antibonding spin-down states are above the conduction bands. Under the trigonal pyramidal crystal field, the degeneracy of the Fe $d$-orbitals is lifted, giving rise to three groups of orbitals: the doubly degenerate states with IRs of $e''$ and $e'$, respectively, and the higher energy $a'$ state.

Knowledge of the IRs facilitates our analysis of intra-defect optical transitions. Many-body excited states are constructed by considering the transfer of an electron from the highest-energy occupied defect states to the lowest-energy unoccupied defect states.[47,48] The rationale for the IR assignments is provided in SI, Note S2.

The optical transition between two states is allowed if $\Gamma_i \otimes \Gamma_\mu \otimes \Gamma_f \supset \Gamma_1$, where $\Gamma_i$ and $\Gamma_f$ are, respectively, the IRs of the initial state and the final state, and $\Gamma_1$ is the fully symmetric IR.[49] $\Gamma_\mu$ refers to the IR of the optical transition dipole moment (TDM) operator, $\boldsymbol{\mu}$. Symmetry-allowed radiative transitions couple spin sublevels with the same spin projection. The difference in frequencies between the ground state $|\pm\rangle$ to excited state $|\pm\rangle$ transitions, relative to that for the $|0\rangle$ state, is about 3.4 THz. This enhances the optical readout accuracy of the spin state by frequency-dependent photoluminescence spectra.

The zero-phonon line (ZPL) transition energy calculated with the HSE06 functional is predicted to be ~ 0.9 eV, which is within the range of telecom wavelengths. While most of the quantities presented in this work can be accurately predicted using DFT, the constrained DFT approach to computing the ZPL transition energy does not account for the possibility of a superposition of excitations for the many-body exciton state. However, for these low-energy transitions, it is expected that the transitions considered here can predominantly account for the many-body exciton state. As DFT (even with the HSE06 functional) underestimates the fundamental gap in these materials, it is possible that the ZPL energy is underestimated by up to a few tenths of eVs. Within the limit of our accuracy, it is reasonable to conclude that the ZPL energy is in the near-IR range.

For $Fe^0$-$MoS_2$, the spin-singlet state lies between the spin-triplet ground state and its first excited state. As such, ISC may be possible between the triplet and singlet states. Applying the condition of $\Gamma_i \otimes \Gamma_{SOC} \otimes \Gamma_f \supset \Gamma_1$ for non-radiative ISC mediated by SOC, and noting that $\Gamma_{SOC}$ is fully-symmetric,[50] ISC transitions can be deduced to be allowed between spin sublevels that belong to IRs having the same characters under symmetry operations common to the respective point groups in both systems. The singlet state has a slight JT distortion towards $C_{2v}$ symmetry, and its IR corresponds to the fully symmetric $A_1$ IR. $C_{2v}$ is a subgroup of $D_{3h}$ (see Figure 1c) with lowered symmetry. Considering the symmetry operations in $C_{2v}$, which are common to both $C_{2v}$ and $D_{3h}$, it can be deduced that symmetry-allowed SOC-mediated ISC can take place between the singlet $A_1$ state and the $A_1'$ and $E'$ sublevels in the triplet state. Consequently,

SOC-mediated ISC can preferentially couple the spin-singlet state with the $E'$ sublevels of the first-excited triplet state and the $A_1'$ sublevel of the ground state in the spin-triplet manifold. Additional ISC mechanisms such as electron-phonon coupling can potentially lead to higher-order transitions that couple other spin-sublevels, especially at higher temperatures.

Because the $E_{xy}'$ sublevels of the first-excited triplet state are lower in energy than the corresponding $A_2''$ sublevel, the preferential coupling of the $E_{xy}'$ sublevels to the spin-singlet state implies that multiple cycles of optical pumping can result in a population inversion within the spin sublevels in the first excited triplet state. As discussed below, the modulus square of TDM, $|\boldsymbol{\mu}|^2$, for the optical transition between the ground and excited triplet states is rather small (3.4 $D^2$). As a result, the system can potentially emit THz photons via magnetic-dipole transitions between the $E_{xy}'$ and $A_2''$ sublevels in the first-excited state, if the lifetime of the first excited triplet state, relative to the ground state, is sufficiently long.

The proposed mechanism for THz emission can be generalized to other potential TM defect systems with a THz ZFS, especially if the population inversion can be achieved in the ground state. Notably, this emission mechanism operates at zero magnetic field, offering a key advantage over systems such as the $NV^-$ centre, where THz emission has been demonstrated only under the application of large external magnetic fields.[51] The zero-field operation presents a promising route toward realising single photon THz emitters,[42,52] which have recently attracted interest for quantum telecommunication and quantum sensing applications.[11]

**$Mn^{1-}$-$WSe_2$.** $Mn^{1-}$-$WSe_2$ also exhibits $D_{3h}$ symmetry, and the analysis is similar to that of $Fe^0$-$MoS_2$, although details differ (see Figure 5). In particular, in the case of $Mn^{1-}$-$WSe_2$, the metastable spin-singlet state is 0.76 eV higher in energy than the first excited state with triplet spin. The fact that there is no intermediate state is highly favourable for high fidelity operations of the qubit through resonant optical excitations with no ISC to a singlet state. In particular, this system is able to leverage on a significant SOC effect for a large ZFS of 1.2 THz while still having a well-defined triplet state with no non-radiative ISC coupling.

**$Mn^{1-}$-$MoS_2$.** As shown in Figure 6, with full-relaxation, the local geometry of $Mn^{1-}$-$MoS_2$ exhibits a JT-distorted $C_{2v}$ symmetry with the in-plane $y$ axis as the two-fold rotation axis (Figure 1b). The Mn dopant has a slight in-plane shift away from one pair of S atoms, so that the Mn-S bond lengths for this pair are 2.34 Å, while the other Mn-S bond lengths are 2.32 Å. The two occupied gap states $b_2$ and $1a_1$ and the lowest unoccupied $2a_1$ state contributed from Mn $d$-orbitals hybridize with adjacent Mo $d$-orbitals and S $p$-orbitals.

The spin-triplet ground state of $Mn^{1-}$-$MoS_2$ ($^3B_2$) consists of the spin sublevels $|-\rangle$ ($B_1$), $|+\rangle$ ($A_2$) and $|0\rangle$ ($A_1$) with energy splittings of up to 0.04 THz and 0.05 THz, respectively (see Figure 6d). The first excited state ($^3B_2$), has a ZFS of around 0.03 THz. There is a reordering of the spin states in the excited state, which contributes to a larger ZPL frequency contrast. The ZPL frequency contrasts for the transitions for $A_1$ and $B_1$ relative to that for $A_2$ are calculated to be 0.09 THz and 0.07 THz, respectively. This contrast is relatively smaller than the $Fe^0$-$MoS_2$ and $Mn^{1-}$-$WSe_2$ systems that have larger ZFS values. However, the spin-sublevels are still more separated compared to GHz qubits (e.g. the $NV^-$ centre in diamond has a ZFS of 2.87 GHz,[4] or 0.003 THz, one order of magnitude smaller in value than the ZFS in $Mn^{1-}$-$MoS_2$). Similar to $Mn^{1-}$-$WSe_2$, the metastable singlet state is higher in energy than the first excited triplet state by 0.36 eV, meaning that there is no intermediate state for ISC transition. Thus, $Mn^{1-}$-$MoS_2$ is also a candidate for high fidelity spin qubits operated by resonant excitations. In this case, the ZPL energy is about 0.6 eV at the DFT (HSE06) level.

## DISCUSSIONS

The results presented here, for 50 different substitutional TM defects in $MoS_2$ and $WSe_2$ MLs, serve as a useful repository of information on the charge and spin state stability of these defect systems, while the specific results on selected spin qubit candidates suggest their potential as high-fidelity qubits operating at higher temperatures. Below we discuss the feasibility of realising the proposed defect systems and their applications in THz technologies.

**Defect creation.** While extrinsic elements can be present as adsorbed or intercalated atoms in 2D materials,[53] Mn and Fe atoms specifically have been experimentally observed to be substitutional cation-site dopants in $MoS_2$.[54,55] In addition, there have been significant recent advancements in the development of scalable approaches to controllably incorporate a range of substitutional point defects in 2D TMDs.[25–27] The advantage of using semiconducting 2D TMDs (compared to materials with larger band gaps) is the relative ease with which the Fermi level can be controlled, for example, using metal contacts,[56] gate control,[57] chemical doping,[58] and proton irradiation.[59] As described above, the respective charge states are accessible for a significant range of Fermi levels in the band gap. The defect formation energies are computed to be 0.5, 0.8 and 2.1 eV, respectively, for neutral $Mn_{Mo}$-$MoS_2$, $Fe_{Mo}$-$MoS_2$ and $Mn_{Mo}$-$WSe_2$ (see SI, Figure S1). The small but positive values indicate the relative ease at which these defects can be created.

**Optical brightness and coupling to light.** While localized TM centres in molecules can have dim or dark optical transitions due to the localized $d$ orbitals in the $d$-$d$ transitions, for our solid-state spin qubits, the orbitals couple to the lattice. The modulus square of the TDM, $|\boldsymbol{\mu}|^2$, is explicitly computed to be 15.2 $D^2$, 4.0 $D^2$ and 3.4 $D^2$ for $Mn^{1-}$-$MoS_2$, $Mn^{1-}$-$WSe_2$ and $Fe^0$-$MoS_2$, respectively. The value for $Mn^{1-}$-$MoS_2$ is reasonably large, while those for $Mn^{1-}$-$WSe_2$ and $Fe^0$-$MoS_2$ are relatively small. To enhance the light-matter interaction, one can also consider using a nanophotonic cavity, both for the ZPL optical pumping and the proposed THz emission. For example, photonic cavities were used to detect microwave emission using ZFS-split levels in SiC defects,[42,52] and single photon emission from defects.[21,22] In addition, we point out that the high frequency contrast in the photoluminescence spectra is advantageous for the operation fidelity of the qubits.

We also comment that the optical transitions in $Mn^{1-}$-$MoS_2$ occur via in-plane polarised light, while those in $Mn^{1-}$-$WSe_2$ and $Fe^0$-$MoS_2$ require an out-of-plane component in the polarisation. The latter can be achieved through various techniques developed in recent years to study the interaction between layered materials with out-of-plane polarised light; these techniques involve either rotating the light source,[60] or rotating the sample.[61]

**THz technologies.** In our proposed approach, the spin state is optically initialized and read out via infrared or near-infrared transitions (e.g., telecom wavelengths), while THz radiation is used to drive Rabi oscillations between spin sublevels. While many research groups working on solid-state spin qubits are set up primarily for qubits with ZFS in the GHz range, it is important to look beyond immediate applications and recognise the substantial investment in mmWave (i.e. THz) technologies in related fields, by funding agencies, scientists and telecommunication companies.[62] There have been significant advancements in the development of THz sources.[28,63,64] In particular THz electron spin resonance techniques, with frequencies greater than 100 GHz, and operating at high signal-to-noise ratios are now available,[65–67] opening up possibilities for controlling qubits with THz ZFS, which can lead to long-term breakthroughs in the operation-fidelity and scalability of the spin qubits. While most of the technological developments in the THz field are focused on classical THz radiation, there has also been nascent interest in quantum THz radiation, and single photon detection of THz radiation has been explored and demonstrated.[11,68,69] Our proposal for a solid-state THz single-photon emitter aligns with the recent interest in emerging quantum THz technologies.

## CONCLUSIONS

In this study, we perform a high-throughput first-principles investigation of 50 substitutional TM defects in ML $MoS_2$ and $WSe_2$ to identify spin-triplet systems with large ZFS in the THz regime. We show that such large ZFS enables the potential operation of spin qubits at higher temperatures. For selected defects, symmetry analysis and electronic structure calculations confirm the feasibility of resonant optical control, supporting high-fidelity qubit operations. Moreover, we demonstrate that spin-selective transitions in specific systems enable population inversion, highlighting their potential as THz single-photon emitters.

Above all, our results establish a class of solid-state quantum defects in 2D materials with potential applications in quantum THz technologies. We hope that the work presented here will catalyse collaborations between the spin qubit and THz communities, opening up a new field of THz quantum technologies that will eventually enable us to realise spin qubits with high fidelities and longer spin coherence times, at higher temperatures, as well as break new grounds in quantum telecommunications.

## METHODS

**ZFS calculations.** In this work, the spin quantization axis $\gamma$ is along the principal magnetic axis by following the convention that $E_{\mathrm{tot}}^{\mathrm{DFT}}(\gamma\gamma)$ has the largest difference from $E_{\mathrm{tot}}^{\mathrm{DFT}}(\alpha\alpha)$ and $E_{\mathrm{tot}}^{\mathrm{DFT}}(\beta\beta)$, where $\alpha$ and $\beta$ are the other two subsidiary axes. The calculation for the first-order ZFS contributed by spin-spin dipole interactions is implemented in VASP.[70] The second-order SOC-contributed ZFS is inherently related to the magnetic anisotropy energy of the defect with the spin Hamiltonian of [41]

$$\hat{H}_{\mathrm{ZFS}} = D^{\mathrm{SO}}\left(\hat{S}_{\gamma}^{2} - \frac{1}{3}S(S+1)\right) + E^{\mathrm{SO}}\left(\hat{S}_{\alpha}^{2} - \hat{S}_{\beta}^{2}\right) = \hat{D}^{\mathrm{SO}} + \hat{E}^{\mathrm{SO}},$$

which yields, for $S = 1$, $\frac{1}{3}D^{\mathrm{SO}} + E^{\mathrm{SO}}$, $\frac{1}{3}D^{\mathrm{SO}} - E^{\mathrm{SO}}$, and $-\frac{2}{3}D^{\mathrm{SO}}$ for $|+\rangle$, $|-\rangle$ and $|0\rangle$ eigenstates, respectively.

For a single spin, with spin states $|\gamma\rangle \equiv |\uparrow\rangle$ and $|-\gamma\rangle \equiv |\downarrow\rangle$, we have

$$|\alpha\rangle = \frac{1}{\sqrt{2}}(|\uparrow\rangle + |\downarrow\rangle),$$

$$|\beta\rangle = \frac{1}{\sqrt{2}}(|\uparrow\rangle + i|\downarrow\rangle).$$

Thus, for two spins in the $S = 1$ system,

$$|\alpha\alpha\rangle = |\alpha\rangle \otimes |\alpha\rangle = \frac{1}{2}(|\uparrow\uparrow\rangle + |\downarrow\downarrow\rangle) + \frac{1}{2}(|\uparrow\downarrow\rangle + |\downarrow\uparrow\rangle),$$

$$|\beta\beta\rangle = |\beta\rangle \otimes |\beta\rangle = \frac{1}{2}(|\uparrow\uparrow\rangle - |\downarrow\downarrow\rangle) + \frac{1}{2}i(|\uparrow\downarrow\rangle + |\downarrow\uparrow\rangle).$$

Therefore, using the Hamiltonian $\widehat{H}_{\mathrm{ZFS}} + \widehat{H}_{\mathrm{elec}}$, the total energies are:

$$E^{\mathrm{DFT}}_{\mathrm{tot},\alpha\alpha} = -\frac{1}{6}D^{\mathrm{SO}} + \frac{1}{2}E^{\mathrm{SO}} + E_{\mathrm{elec}},$$

$$E^{\mathrm{DFT}}_{\mathrm{tot},\beta\beta} = -\frac{1}{6}D^{\mathrm{SO}} - \frac{1}{2}E^{\mathrm{SO}} + E_{\mathrm{elec}},$$

$$E^{\mathrm{DFT}}_{\mathrm{tot},\gamma\gamma} = \frac{1}{3}D^{\mathrm{SO}} + E_{\mathrm{elec}},$$

where $E_{\mathrm{elec}}$ is introduced as the total electronic energy, which we assume to be independent of the orientation of the spins. We find that for the spin-triplet ground states of our candidates, the $\gamma$-direction is aligned with the $z$-direction, where $x$, $y$, and $z$ are chosen to be the axis directions in Figure 1, which also correspond to the eigenvectors of the spin-spin ZFS Hamiltonian. Now, the value of $D^{SO}$ and $E^{SO}$ can be directly calculated by DFT through:

$$D^{\mathrm{SO}} = 2E^{\mathrm{DFT}}_{\mathrm{tot},zz} - \left[E^{\mathrm{DFT}}_{\mathrm{tot},xx} + E^{\mathrm{DFT}}_{\mathrm{tot},yy}\right],$$

$$E^{\mathrm{SO}} = E^{\mathrm{DFT}}_{\mathrm{tot},xx} - E^{\mathrm{DFT}}_{\mathrm{tot},yy}.$$

**Other computational details**

Other computational details such as DFT computational details, defect formation energies, CTLs, TDMs, singlet-triplet energies, and excited states can be found in the Methods section in the SI where we have also studied the effect of the supercell size on the results.

## ASSOCIATED CONTENT

## Supporting Information

Methods section including DFT computational details, details on calculations for defect formation energies and CTLs, singlet and triplet energies, ZFS, TDM, excited-state calculations, and SOC; Details on the exchange-correlation functional and whether SOC was included in different parts of the calculations; Point group symmetries of TM defects; Thermodynamic CTLs; Singlet and triplet energies for $Mn^{1-}$ and $Fe^{0}$ defects; ZFS values for TM defects; Formation energy diagram of $Mn_{Mo/W}$ and $Fe_{Mo/W}$; ZFS value as a function of in-plane strain; Relaxed structures of $Mn^{1-}$ and $Fe^{0}$ defects; Band structures for $Mn^{1-}$ and $Fe^{0}$ defects; Symmetry-driven defect geometry optimization; and Details on the group theory analysis.

## Author contributions

J.Z. performed the calculations and data analysis under the supervision of S.Y.Q. All authors contributed to the discussion and writing of this paper.

## Notes

The authors declare no competing interests.

## ACKNOWLEDGEMENTS

The project is supported by the Ministry of Education (MOE), Singapore, under grant number MOE2018-T3-1-005. The computational work for this article was partially performed on resources of the National Supercomputing Centre, Singapore (https://www.nscc.sg). Work performed at the Center for Nanoscale Materials, a U.S. Department of Energy Office of Science User Facility, was supported by the U.S. DOE, Office of Basic Energy Sciences, under Contract No. DE-AC02-06CH11357. The authors thank Y. Chen, N.L.Q. Cheng, K. Ulman, L. Loh, G. Eda, W. Huang, and Z. Hu for discussions, and W. He for assistance with the illustration of the Table of Contents graphic.

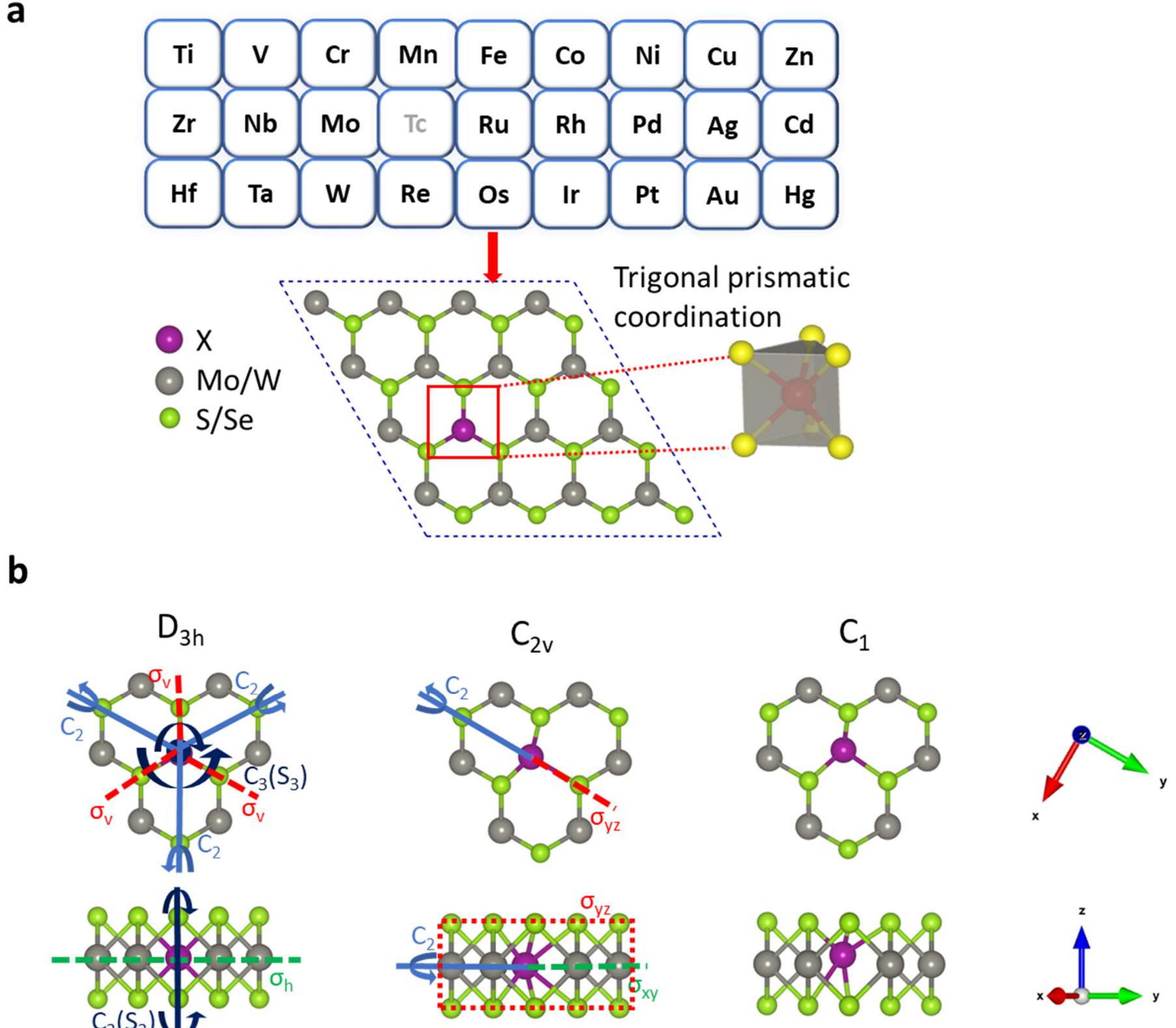

**Figure 1.** (a) Schematic showing the creation of a substitutional defect where one Mo/W atom in a 4×4 supercell of monolayer 2H-$MoS_2$/2H-$WSe_2$ is replaced by TM atom X. The trigonal prismatic coordination shown here represents the initial atomic structure before structural relaxation. (b) Top views (top) and side views (bottom) of the possible local atomic structure and symmetries for the TM substitutional defects with spin-triplet ground states, after structural relaxation.

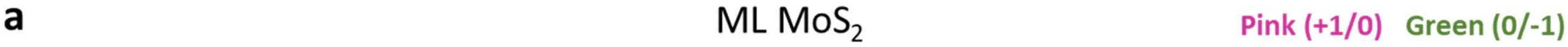

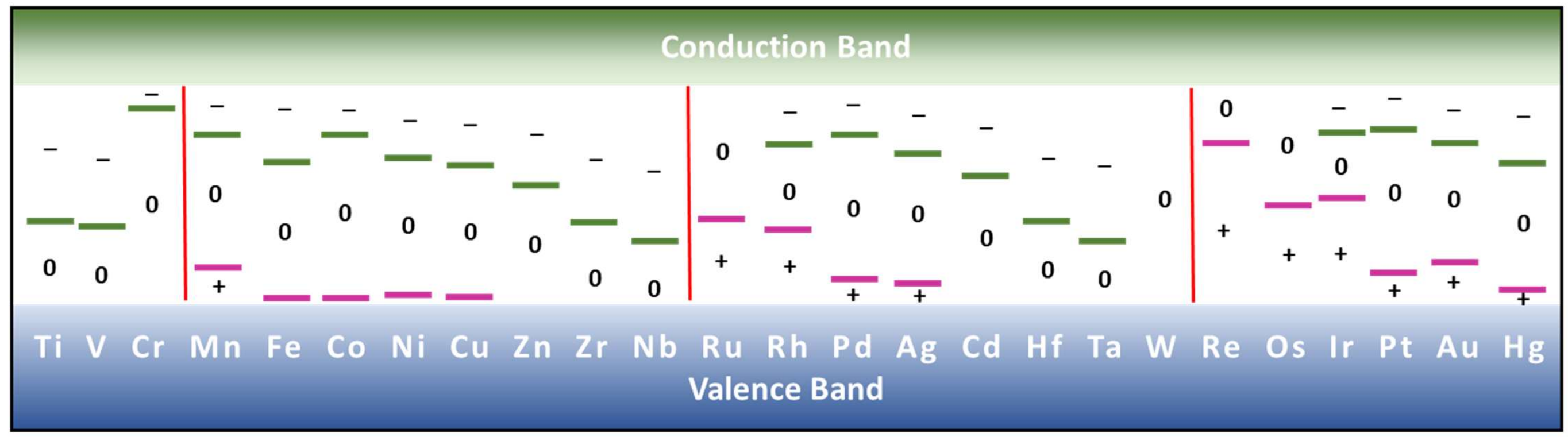

b

ML WSe2

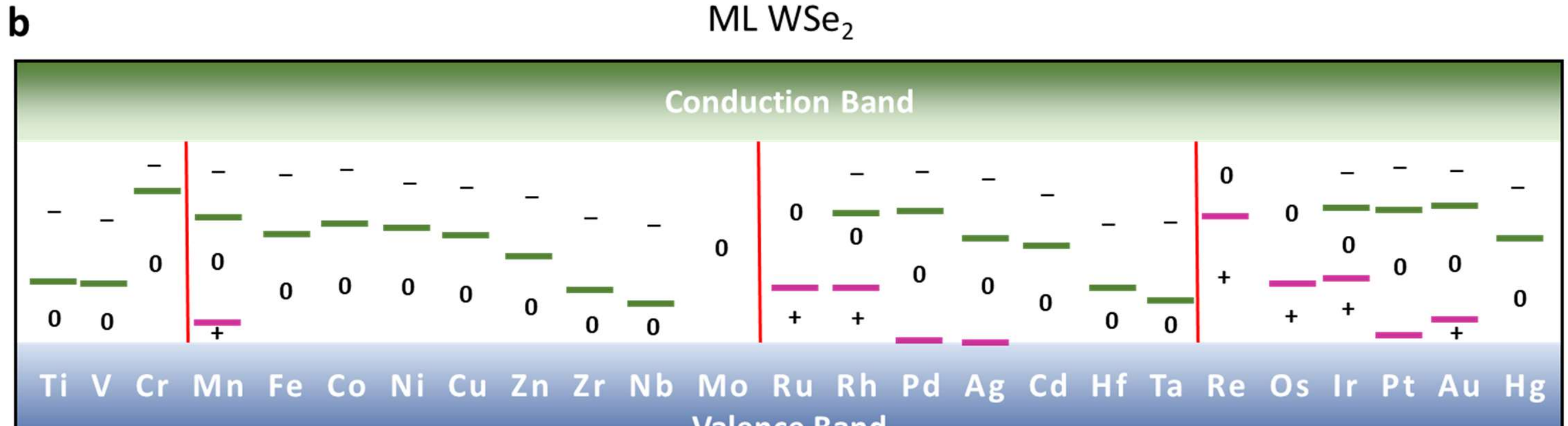

**Figure 2.** Thermodynamic charge transition levels (CTLs) for TM substitutional defects in ML (a) $MoS_2$ and (b) $WSe_2$. The vertical axis is energy, and the CTL energies are aligned with the valence band maximum (VBM; PBE band gaps are used here (1.68 eV for $MoS_2$ and 1.54 eV for $WSe_2$)) of the pristine TMD. (+1/0) levels (purple) and (0/–1) levels (green) denote the transition levels from the charge state +1 to 0, and from 0 to –1, respectively. 0, +, and – indicate the neutral, +1, and –1 charge state within the Fermi level range. Based on calculations on a few defects (see also SI, Figure S1), the higher charge states are unlikely to be stable for Fermi levels within the band gap in these systems. The Group VIB elements (eg. Mo, W, Cr) are marked with a vertical line on their right. See SI, Table S3 for the CTL values.

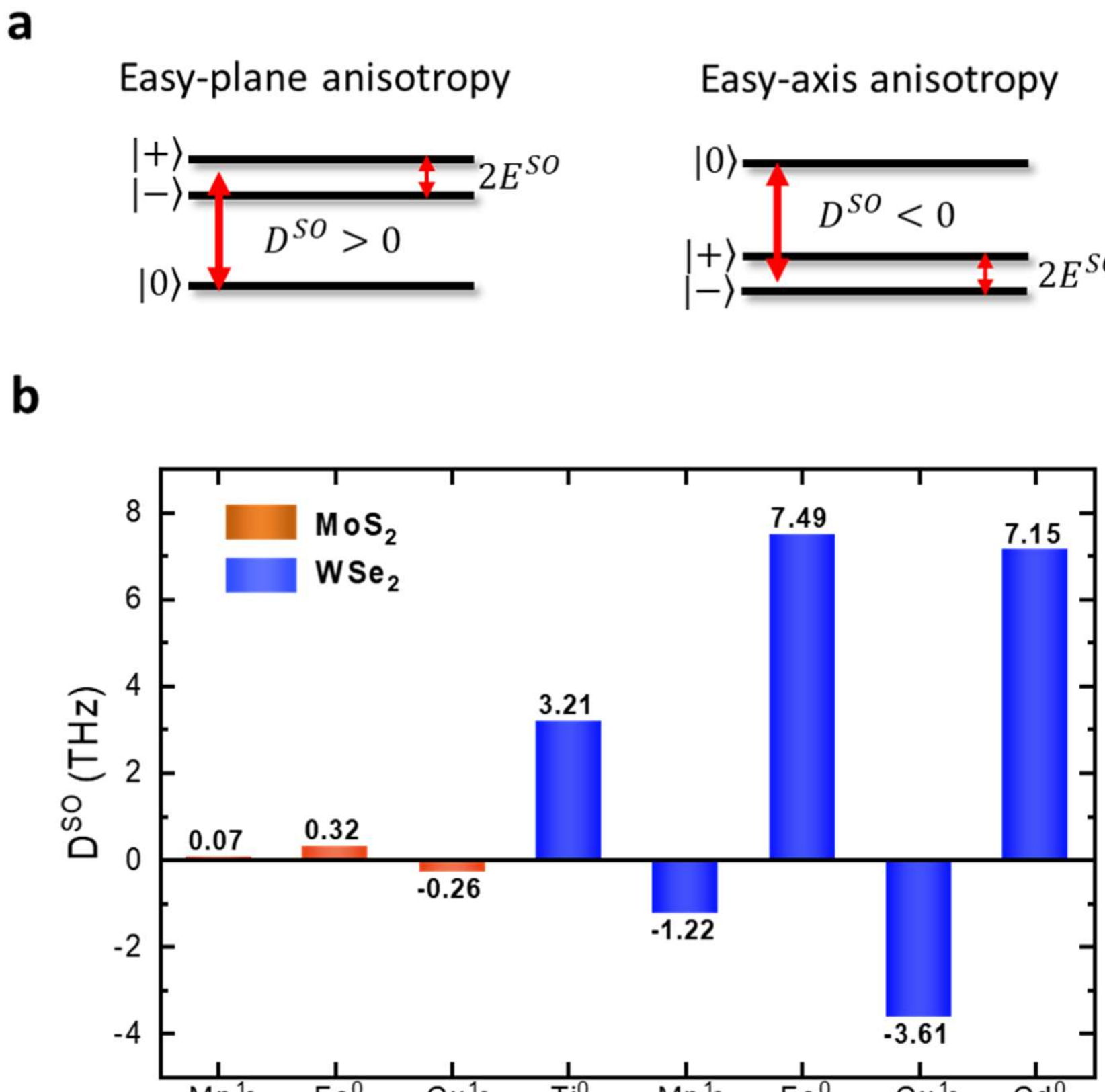

**Figure 3.** (a) Spin sublevel diagram for spin-triplet state for easy-plane and easy-axis magnetic anisotropy for the case where $E^{SO} > 0$. (b) $D^{SO}$ values of ZFS for TM dopants in ML $MoS_2$ and $WSe_2$.

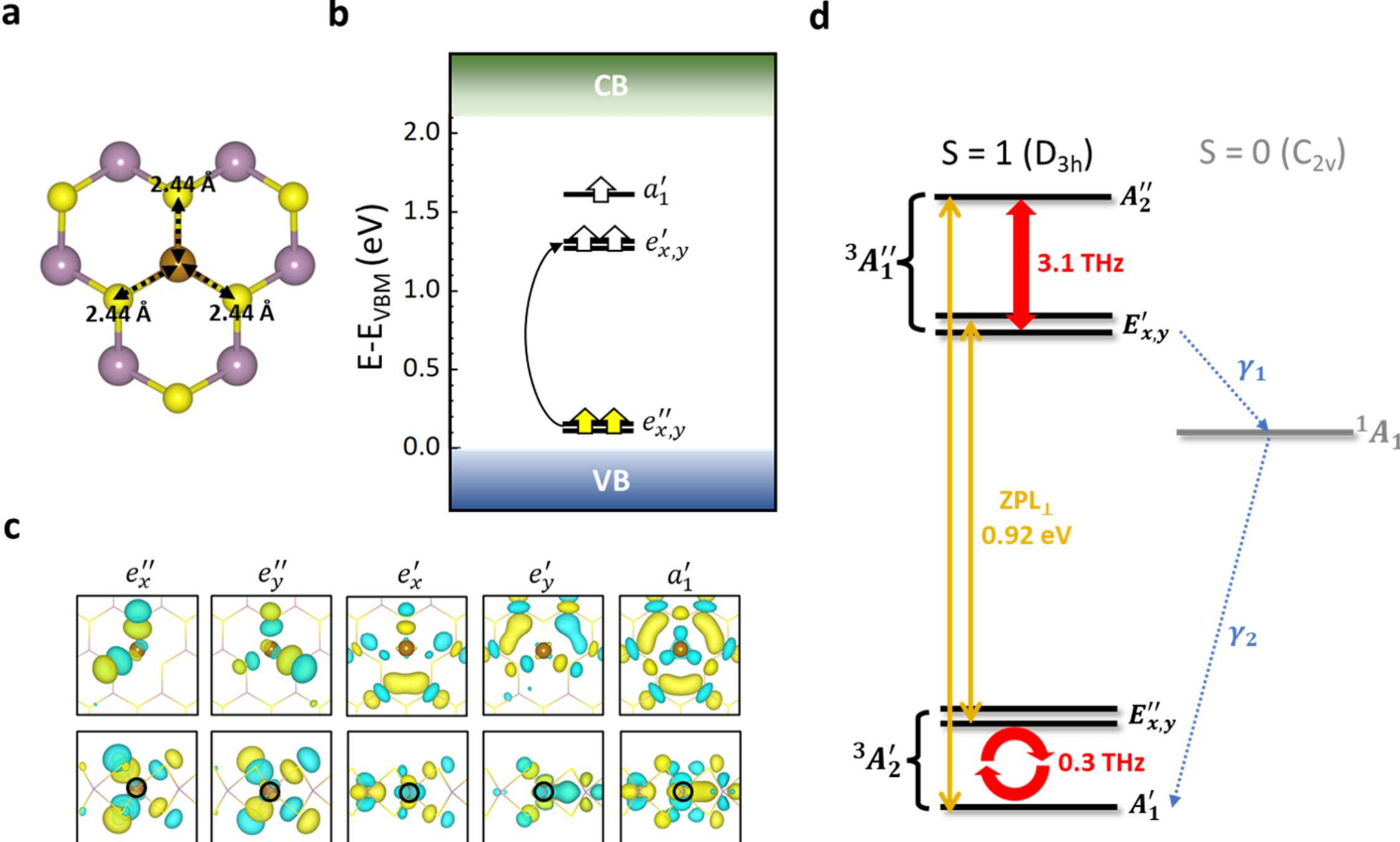

**Figure 4.** (a) Relaxed geometry and (b) single-particle Kohn-Sham defect states of $Fe^0$-$MoS_2$. The yellow solid and hollow arrows mark the occupied and unoccupied in-gap defect states, respectively, which are labelled according to their irreducible representations transforming under group symmetry. The black arrow denotes the electron transfer excitation leading to the many-body excited states in (d). (c) Visualization of the real part of the wavefunctions at the Gamma point in the top-view (top) and side-view (bottom) with an isosurface value of 30% of the maximum. The defect sites are circled in black for the side-view. (d) Many-body level structure with intersystem crossing pathways $\gamma_1$ and $\gamma_2$ (blue), the zero-field splitting of ground and excited states (red), and the ZPL energies of optically-allowed spin sublevel transitions with polarized photons (yellow), indicated. The $x$, $y$ and $z$ directions follow the coordinates indicated in Figure 1.

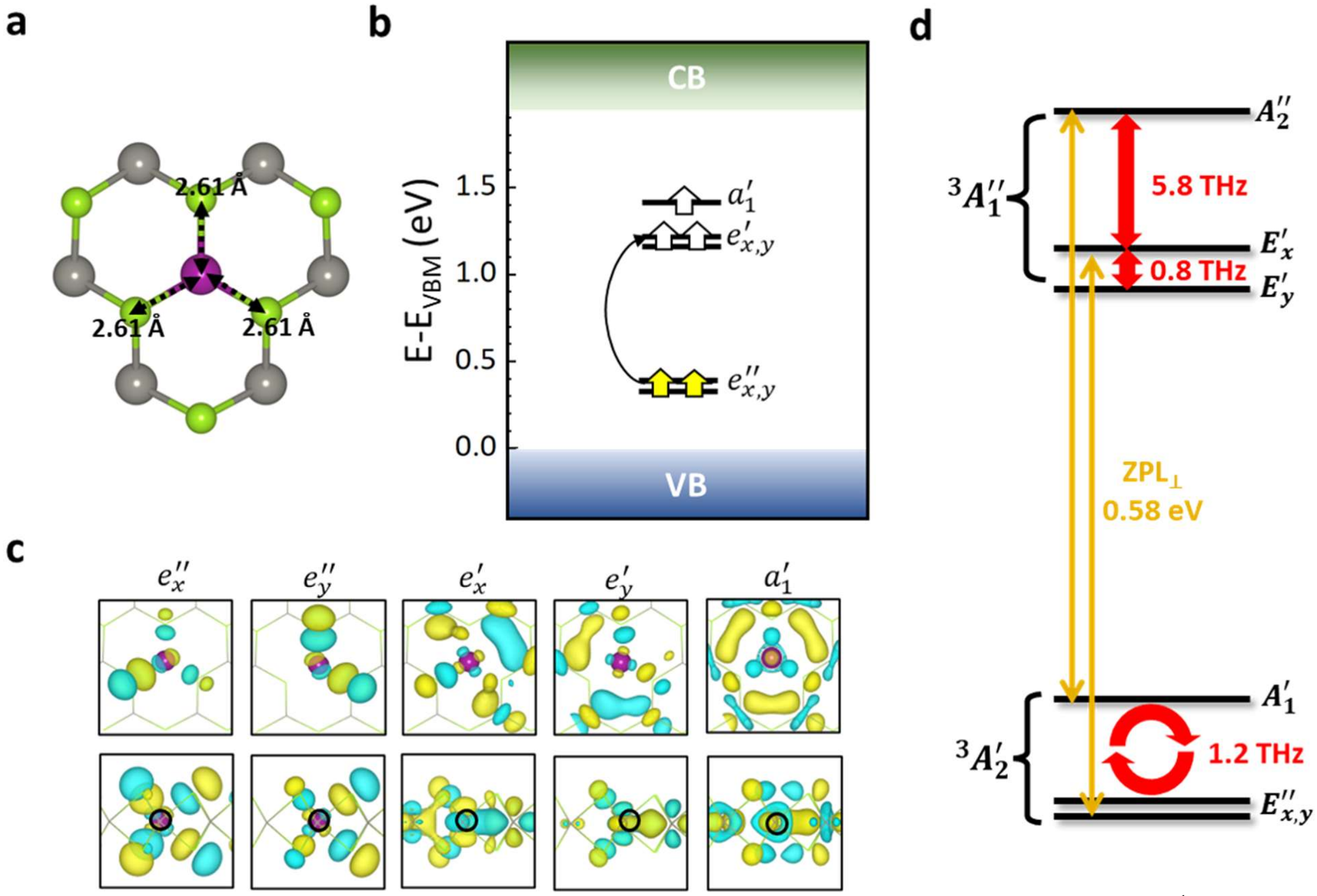

**Figure 5.** (a) Relaxed geometry and (b) single-particle Kohn-Sham defect states of $Mn^{1-}$-$WSe_2$. The yellow solid and hollow arrows mark the occupied and unoccupied in-gap defect states, respectively, which are labelled according to their IRs transforming under group symmetry. The black arrows denote the electron transfer excitations leading to the many-body states in (d). (c) Visualization of the real part of the wavefunctions at the Gamma point in the top-view (top) and side-view (bottom) with an isosurface value of 30% of the maximum. The defect sites are circled in black for side-view. (d) Many-body level structure with the zero-field splitting of ground and excited states (red), and the ZPL energies of optically-allowed spin sublevel transitions with polarized photons (yellow), indicated. The $x$, $y$ and $z$ directions follow the coordinates indicated in Figure 1.

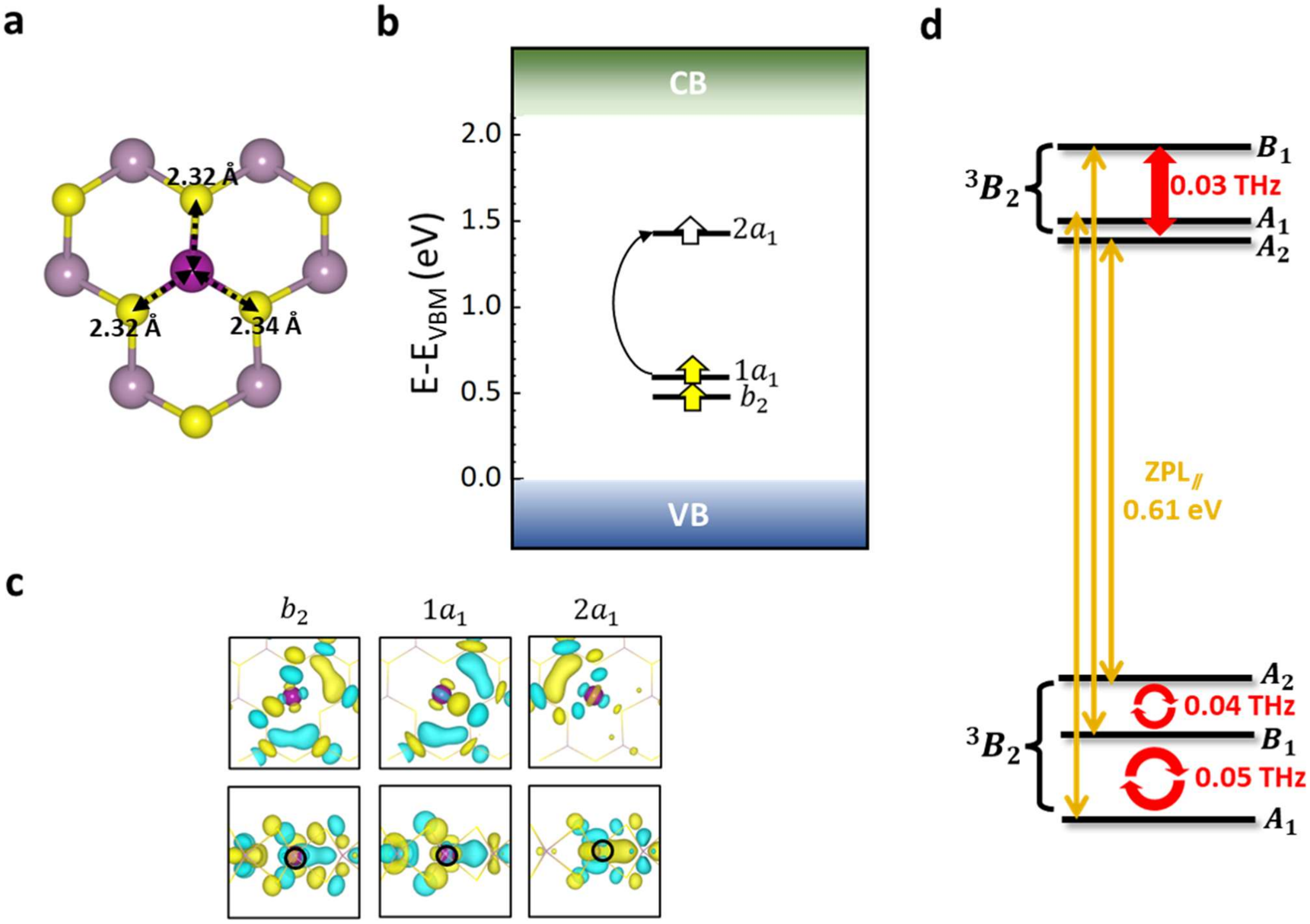

**Figure 6.** (a) Relaxed geometry and (b) single-particle Kohn-Sham defect states of $Mn^{1-}$-$MoS_2$. The yellow solid and hollow arrows mark the occupied and unoccupied in-gap defect states, respectively, which are labelled according to their IRs transforming under group symmetry. The black arrow denotes the electron transfer excitation leading to the many-body excited states in (d). (c) Visualization of the real part of the wavefunctions at the Gamma point in the top-view (top) and side-view (bottom) with an isosurface value of 30% of the maximum. The defect sites are circled in black for the side-view. (d) Many-body level structure with the zero-field splitting of ground and excited states (red), and ZPL energies of optically-allowed spin sublevel transitions with polarized photons (yellow), indicated. The $x$, $y$ and $z$ directions follow the coordinates indicated in Figure 1 and the corresponding character table is shown in SI, Note S2(iii).

| TM defect | Charge state | Symmetry | $\Delta E$ (Singlet-Triplet) (eV) |
|---|---|---|---|
| **Host - $MoS_2$** | | | |
| Mn | -1 | $C_{2v}$ | 0.97 (0.80) |
| Fe | 0 | $D_{3h}$ | 0.85 (0.43) |
| Cu | -1 | $C_1$ | 0.22 (0.01) |
| **Host - $WSe_2$** | | | |
| Ti | 0 | $C_{2v}$ | 0.01 (0.02) |
| Mn | -1 | $D_{3h}$ | 1.34 (1.04) |
| Fe | 0 | $D_{3h}$ | 1.09 (0.65) |
| Cu | -1 | $C_1$ | 0.08 (0.03) |
| Cd | 0 | $C_{2v}$ | 0.01 (0.02) |

**Table 1.** Singlet-triplet energy differences $\Delta E$ of TM substitutional defects in ML TMDs calculated by the HSE06 functional. The numbers in bracket are $\Delta E$ obtained by the PBE+U functional. Positive values indicate that the spin-triplet state is more stable. Point group symmetries of the spin-triplet state are identified by the local geometry around the defect and by visualization of defect states, in the geometry obtained following relaxation with the HSE06 functional. For $Mn^{1-}$ and $Fe^{0}$ defects, spin-purification correction is required and is applied to the singlet energy; see SI, Table S4 for details.

*Supporting Information*

# Quantum Defects in 2D Transition Metal Dichalcogenides for Terahertz Technologies

Jingda Zhang[1] and Su Ying Quek[1,2,3,4,5*]

[1]Department of Physics, National University of Singapore, Singapore 117551

[2]Centre for Advanced 2D Materials, National University of Singapore, Singapore 117546

[3]Department of Materials Science and Engineering, National University of Singapore, Singapore 117575

[4]Integrative Sciences and Engineering Programme, NUS Graduate School, National University of Singapore, Singapore 119077

[5]NUS College, National University of Singapore, Singapore 138593

*Corresponding author: S.Y.Q. (email: phyqsy@nus.edu.sg)

# METHODS

**Density functional theory computational details.** Our spin-polarized DFT calculations are performed using VASP[1] within the projector-augmented wave[2] formalism and a plane-wave energy cutoff of 450 eV. For the preliminary screening calculations for appropriate charge states and spin-triplet ground states, we adopt the generalized gradient approximation to the exchange-correlation functional, as parameterized by Perdew, Burke and Ernzerhof (PBE)[3] with an on-site Hubbard U of 3 eV for all 3d elements.[4] The hybrid Heyd-Scuserial-Ernzerhof (HSE06)[5,6] functional is used for defect structure relaxations, calculations of zero-field splitting (ZFS), excited-state properties and formation energy diagrams of selected candidates. For the defect structure relaxations in the screening process, a slight tilted out-of-plane displacement is introduced on the initial dopant position (See Note S1). To determine the defect local symmetry, a precision threshold of 0.005 Å for the bond lengths between the dopant and the adjacent chalcogen atoms is applied, ensuring that the reported bond lengths are precise to the second decimal place. We use a 4×4×1 supercell, with the Brillouin zone sampled by a 3×3×1 Monkhorst-Pack k-mesh. We have also checked the atomic structures, ZFS and electronic structures for key results with a 6×6×1 supercell using a 2×2×1 k-mesh (see Table S5 and accompanying discussion, Figure S3-S6). A 15-Å-thick vacuum layer is included to avoid image interaction in the $z$-direction. The structures are relaxed until the magnitudes of the forces are less than 0.01 eV/ Å for PBE and 0.02 eV/ Å for HSE06, respectively.

**Defect formation and charge transition levels.** The defect formation energy is computed following:[7]

$$E^f\left[TM^q_{\mathrm{Mo/W}}\right] = E_{\mathrm{tot}}\left[TM^q_{\mathrm{Mo/W}}\right] - E_{\mathrm{tot}}[\mathrm{Pristine}] + \mu_{\mathrm{Mo/W}} - \mu_{TM} + q(\varepsilon_F + E_{\mathrm{VBM}}) + \Delta_q.$$

$TM$ stands for transition metal impurities. $E_{\mathrm{tot}}$ is the total energy of the defect and pristine system. $\mu$ is the chemical potential for the element species. Comments on the chemical potentials used are provided in the discussion accompanying Figure S1. $\varepsilon_F$ is the Fermi level referenced to $E_{\mathrm{VBM}}$, the VBM of the pristine cell. $\Delta_q$ represents the charge-dependent finite-size correction by Freysoldt-Neugebauer-Van de Walle (FNV)[8] calculated using the sxdefectalign2d tool[9] to eliminate the artificial electrostatic interaction between the periodic defect images and background charge. The charge transition levels (CTLs) can therefore be computed as:

$$\varepsilon_F(q+1/q) = \left(E_{tot}[TM^{q+1}] + \Delta_{q+1}\right) - \left(E_{tot}[TM^q] + \Delta_q\right) - E_{\mathrm{VBM}}.$$

**Singlet and triplet energies.** To determine the singlet–triplet energy differences (ΔE), we compute the total energies using spin-polarized calculations for both spin-triplet and spin-singlet configurations by constraining total magnetic moments to be 0 $\mu_B$ and 2 $\mu_B$, respectively. For the spin-singlet state, we identify cases where spin-unrestricted (spin-polarized) calculations yield broken-symmetry solutions with residual local magnetic moments, indicating possible spin contamination. In such cases, we perform spin-restricted (non-spin-polarized) calculations to approximate a closed-shell singlet state.[10] The results are summarized in Table S4.

**Zero-field splitting.** As derived in the main text, the value of $D^{\mathrm{SO}}$ and $E^{\mathrm{SO}}$ can be directly calculated by DFT through:

$$D^{\mathrm{SO}} = 2E^{\mathrm{DFT}}_{\mathrm{tot},zz} - \left[E^{\mathrm{DFT}}_{\mathrm{tot},xx} + E^{\mathrm{DFT}}_{\mathrm{tot},yy}\right],$$

$$E^{\mathrm{SO}} = E^{\mathrm{DFT}}_{\mathrm{tot},xx} - E^{\mathrm{DFT}}_{\mathrm{tot},yy}.$$

For the calculation of total energies, we first carry out a self-consistent collinear calculation without SOC. Reading from the ground-state charge density of the system, the SOC can be treated as a perturbation in non-self-consistent calculations (see e.g., Eq. (6) of Ref. 11)). The total energy of the system is calculated with a convergence threshold of $10^{-7}$ eV for different spin orientations.

**Transition dipole moment.** The transition dipole moment (TDM) is a complex vector quantity describing the transition for a system interacting with an electromagnetic wave. For optical spectra, i.e., the electric-dipole transition type of intra-defect transitions, the TDM, $\boldsymbol{\mu}$ between an initial state $a$ and a final state $b$ is evaluated based on single-particle wavefunction by HSE06:

$$\boldsymbol{\mu} = \frac{i\hbar}{(E_b - E_a)m_e}\langle\psi_b|\boldsymbol{p}|\psi_a\rangle = \frac{\hbar^2}{(E_b - E_a)m_e}\sum_i C_{ai}C^*_{bi}\boldsymbol{G}_i,$$

where $\psi_a$ and $\psi_b$ are energy eigenstates with energy $E_a$ and $E_b$; $C_{ai}$, $C^*_{bi}$, and $\boldsymbol{G}_i$ are plane-wave coefficients and reciprocal space vector with the same $\boldsymbol{k}$ vector, respectively, summed over the plane-waves. Its direction gives the polarization of the transition, while $|\boldsymbol{\mu}|^2$ reflects the transition probability.

**Excited-state calculations.** The excited-state total energies are computed by the ΔSCF method[12] using HSE06 for constraining the occupations of the Kohn-Sham orbitals for the desired electronic configuration of excited states.

**Spin-orbit coupling.** In Table S1 below, we provide an overview of when and how spin-orbit coupling (SOC) is included in the calculation, together with the exchange-correlation functional used.

**Table S1.** This table shows the exchange-correlation functional used in the final results of different types of calculations presented in the manuscript, together with information on whether and how SOC was included in the calculation.

| Calculations | Functional | SOC Included |
| --- | --- | --- |
| Geometry and charge states screening | PBE+U | None |
| Singlet-triplet energy gap | PBE+U and HSE06 | None |
| Magnetic anisotropy calculations for ZFS | HSE06 | Non-self-consistently |
| Spin-state stability | PBE+U and HSE06 | Self-consistently |
| Structural relaxation for candidates | HSE06 | None |
| ZPL transitions | HSE06 | None |
| Band structures | HSE06 | Non-self-consistently |
| Formation energy diagram | HSE06 | Self-consistently |
| Transition dipole moments | HSE06 | Self-consistently |

**Table S2.** Point group symmetries of TM substitutional defects with different charge states in ML $MoS_2$ and $WSe_2$ obtained by the high-throughput screening process, with geometry relaxation starting from a $C_1$ geometry for the defect (see Note S1).

| Host | $MoS_2$ | | | $WSe_2$ | | |
|---|---|---|---|---|---|---|
| **Charge** | **-1** | **0** | **+1** | **-1** | **0** | **+1** |
| Ti | $D_{3h}$ | $D_{3h}$ | $D_{3h}$ | $D_{3h}$ | $C_{2v}$ | $D_{3h}$ |
| V | $D_{3h}$ | $D_{3h}$ | $D_{3h}$ | $D_{3h}$ | $D_{3h}$ | $D_{3h}$ |
| Cr | $C_{2v}$ | $D_{3h}$ | $D_{3h}$ | $C_{2v}$ | $D_{3h}$ | $D_{3h}$ |
| Mn | $C_{2v}$ | $C_{2v}$ | $D_{3h}$ | $C_h$ | $C_{2v}$ | $D_{3h}$ |
| Fe | $D_{3h}$ | $D_{3h}$ | $C_{2v}$ | $C_{2v}$ | $D_{3h}$ | $C_{2v}$ |
| Co | $C_1$ | $C_1$ | $D_{3h}$ | $C_1$ | $C_1$ | $D_{3h}$ |
| Ni | $C_s$ | $C_s$ | $C_s$ | $C_s$ | $C_s$ | $C_s$ |
| Cu | $C_s$ | $C_s$ | $C_s$ | $C_s$ | $C_s$ | $C_s$ |
| Zn | $C_{2v}$ | $D_{3h}$ | $C_1$ | $C_{2v}$ | $D_{3h}$ | $C_1$ |
| Zr | $D_{3h}$ | $D_{3h}$ | $D_{3h}$ | $D_{3h}$ | $D_{3h}$ | $D_{3h}$ |
| Nb | $D_{3h}$ | $D_{3h}$ | $D_{3h}$ | $D_{3h}$ | $D_{3h}$ | $D_{3h}$ |
| Ru | $C_{2v}$ | $C_{2v}$ | $C_{2v}$ | $C_{2v}$ | $C_{2v}$ | $C_{2v}$ |
| Rh | $C_1$ | $C_1$ | $C_{2v}$ | $C_1$ | $C_1$ | $C_{2v}$ |
| Pd | $C_s$ | $C_s$ | $C_1$ | $C_s$ | $C_s$ | $C_1$ |
| Ag | $C_s$ | $C_1$ | $C_{3v}$ | $C_s$ | $C_1$ | $C_1$ |
| Cd | $C_{2v}$ | $D_{3h}$ | $C_s$ | $D_{3h}$ | $D_{3h}$ | $C_h$ |
| Hf | $D_{3h}$ | $D_{3h}$ | $D_{3h}$ | $D_{3h}$ | $D_{3h}$ | $D_{3h}$ |
| Ta | $D_{3h}$ | $D_{3h}$ | $D_{3h}$ | $D_{3h}$ | $D_{3h}$ | $D_{3h}$ |
| Re | $C_{2v}$ | $C_{2v}$ | $D_{3h}$ | $C_{2v}$ | $C_{2v}$ | $D_{3h}$ |
| Os | $C_h$ | $C_{2v}$ | $C_{2v}$ | $C_h$ | $C_{2v}$ | $C_{2v}$ |
| Ir | $C_1$ | $C_1$ | $C_{2v}$ | $C_1$ | $C_h$ | $C_{2v}$ |
| Pt | $C_s$ | $C_1$ | $C_1$ | $C_s$ | $C_1$ | $C_1$ |
| Au | $C_s$ | $C_s$ | $C_s$ | $C_s$ | $C_s$ | $C_s$ |
| Hg | $C_s$ | $C_{3v}$ | $C_s$ | $D_{3h}$ | $D_{3h}$ | $C_s$ |
| W; Mo | $D_{3h}$ | $D_{3h}$ | $D_{3h}$ | $D_{3h}$ | $D_{3h}$ | $D_{3h}$ |

**Table S3.** Thermodynamic CTLs for TM substitutional defects in ML $MoS_2$ and $WSe_2$ (using PBE+U, U = 3 eV for 3d TM dopants; the band gap of the pristine ML $MoS_2$ and ML $WSe_2$ are calculated to be 1.68 eV and 1.54 eV, respectively.).

| Host | $MoS_2$ | | $WSe_2$ | |
|---|---|---|---|---|
| CTLs (eV) | (+1/0) | (0/-1) | (+1/0) | (0/-1) |
| Ti | -0.39 | 0.70 | -0.41 | 0.47 |
| V | -0.43 | 0.65 | -0.39 | 0.48 |
| Cr | -0.30 | 1.48 | -0.35 | 1.17 |
| Mn | 0.30 | 1.33 | 0.16 | 0.99 |
| Fe | 0.07 | 1.09 | -0.08 | 0.88 |
| Co | 0.07 | 1.31 | -0.02 | 1.03 |
| Ni | 0.06 | 1.18 | -0.22 | 0.86 |
| Cu | 0.04 | 1.08 | -0.03 | 0.87 |
| Zn | -0.10 | 0.92 | -0.17 | 0.67 |
| Zr | -0.42 | 0.65 | -0.45 | 0.40 |
| Nb | -0.39 | 0.51 | -0.53 | 0.29 |
| Ru | 0.63 | 1.88 | 0.42 | 1.55 |
| Rh | 0.57 | 1.25 | 0.43 | 0.98 |
| Pd | 0.18 | 1.32 | 0.04 | 1.03 |
| Ag | 0.16 | 1.12 | 0.04 | 0.83 |
| Cd | -0.03 | 1.01 | -0.12 | 0.78 |
| Hf | -0.41 | 0.66 | -0.44 | 0.42 |
| Ta | -0.38 | 0.52 | -0.51 | 0.31 |
| Re | 1.24 | 2.05 | 0.97 | 1.60 |
| Os | 0.73 | 2.02 | 0.48 | 1.68 |
| Ir | 0.79 | 1.30 | 0.55 | 1.06 |
| Pt | 0.22 | 1.34 | 0.05 | 1.04 |
| Au | 0.33 | 1.30 | 0.18 | 1.07 |
| Hg | 0.11 | 1.11 | -0.02 | 0.84 |
| W; Mo | -0.44 | 2.30 | -0.57 | 1.98 |

**Table S4.** Singlet and triplet energies for the $Mn^{1-}$ and $Fe^{0}$ defects. Energies are in eV.

| **Functional** | **PBE+U** | | | **HSE06** | | |
|---|---|---|---|---|---|---|
| Spin state | Singlet | Singlet | Triplet | Singlet | Singlet | Triplet |
| Spin restriction | Non-polarized | Polarized | Polarized | Non-polarized | Polarized | Polarized |
| $Mn^{1-}$-$MoS_2$ | -342.30 | -342.50 | -343.10 | -416.46 | -416.73 | -417.43 |
| $Mn^{1-}$-$WSe_2$ | -337.20 | -337.43 | -338.24 | -408.57 | -409.08 | -409.91 |
| $Fe^{0}$-$MoS_2$ | -340.46 | -340.49 | -340.89 | -414.12 | -414.28 | -414.98 |
| $Fe^{0}$-$WSe_2$ | -335.61 | -335.73 | -336.25 | -406.49 | -406.97 | -407.59 |

**Table S5.** Zero-field splitting (ZFS) values for TM substitutional defects in ML $MoS_2$ and $WSe_2$ (HSE06 including spin-orbit coupling (SOC)). For these values reported here, the out-of-plane direction is the principal axis that also corresponds to the spin quantisation axis intrinsic to the system.

| Defects | Symmetry | $D^{SO}$(GHz) | $D^{SS}$(GHz) | $E^{SO}$(GHz) | $E^{SS}$(GHz) | $D^{tot}$(THz) |
|---|---|---|---|---|---|---|
| | | | **Host - $MoS_2$** | | | |
| $Mn^{1-}$ | $C_{2v}$ | 49.3 | 19.7 | 21.2 | -0.3 | **0.07** |
| $Fe^{0}$ | $D_{3h}$ | 306.4 | 16.3 | -0.5 | 0.0 | **0.32** |
| $Cu^{1-}$ | $C_1$ | -262.2 | 0.1 | 40.7 | -1.7 | **-0.26** |
| | | | **Host - $WSe_2$** | | | |
| $Ti^{0}$ | $C_{2v}$ | 3214.6 | 0.9 | 23.0 | -0.1 | **3.21** |
| $Mn^{1-}$ | $D_{3h}$ | -1243.1 | 24.5 | 104.5 | 0.2 | **-1.22** |
| $Fe^{0}$ | $D_{3h}$ | 7482.4 | 10.7 | 0.9 | 0.1 | **7.49** |
| $Cu^{1-}$ | $C_1$ | -3608.4 | 0.5 | -2.3 | 1.4 | **-3.61** |
| $Cd^{0}$ | $C_{2v}$ | 7147.2 | 0.9 | -3.7 | 0.0 | **7.15** |

We have investigated the finite-size effect[13] on ZFS using PBE+U (U = 3 eV). For $Fe^{0}$-$MoS_2$, the $D^{SO}$ is 0.40 THz for both the 4×4×1 and 6×6×1 supercells. In $Mn^{1-}$-$MoS_2$, the values are –0.28 THz and –0.22 THz, respectively. Given the computational costs associated with introducing SOC in larger supercells, the more cost-effective 4×4×1 supercell offers fairly reasonable results.

Note that, as discussed in Ref. 14, PBE calculations tend to overestimate the SOC contributions to SOC, and it is recommended that HSE06 calculations be performed for the ZFS values, as we have done for the values reported in the main text.

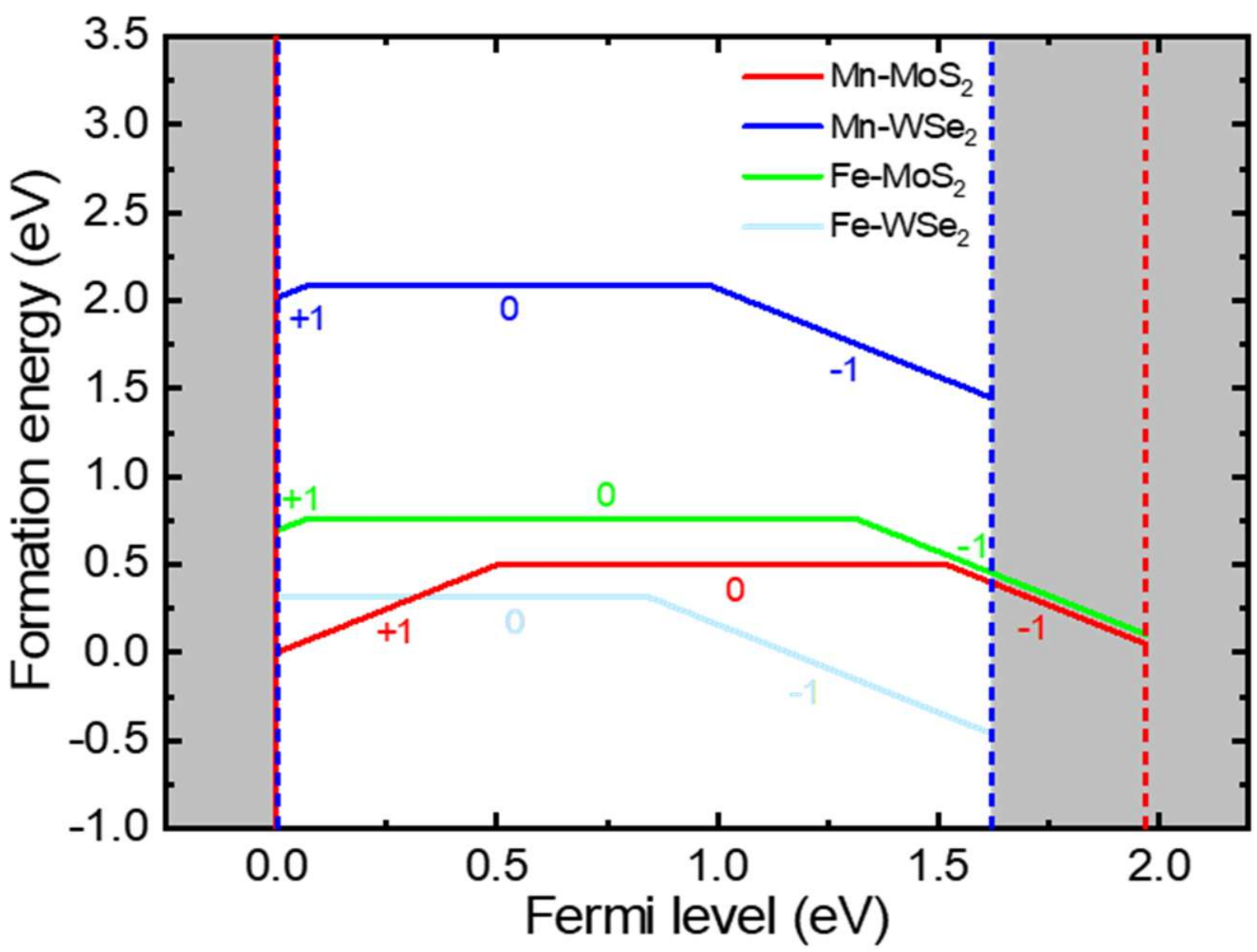

**Figure S1. Formation energy diagram of $Mn_{Mo/W}$, $Fe_{MoW}$ in $MoS_2$ and $WSe_2$** using HSE06 including SOC. The dashed lines indicate the VBM and CBM of ML $MoS_2$ (red) and ML $WSe_2$ (blue). The band gap of pristine ML $MoS_2$ and ML $WSe_2$ are calculated to be 1.97 eV and 1.62 eV, respectively. The VBM of all systems are aligned. Only charge states with the lowest formation energies are shown. The higher charge states (+2, –2, etc.) are found to be unstable for Fermi levels within the band gap.

The formation energy values are determined with chemical potential $\mu$ under constraints:

$$\mu_M + 2\mu_X = E_{\text{tot}}[MX_2],$$

$$\mu_{TM} + \mu_X \le E_{\text{tot}}[(TM)X],$$

$$\mu_M \le E_{\text{tot}}[M],$$

$$\mu_X \le E_{\text{tot}}[X],$$

$$\mu_{TM} \le E_{\text{tot}}[TM].$$

Here, $M$ = Mo/W, $X$ = S/Se, $TM$ = Fe/Mn. Hence, the boundary for the least limit of chemical potential terms can be derived to be:

$$(\mu_M - \mu_{TM})_{min} = E_{\text{tot}}[MX_2] - E_{\text{tot}}[(TM)X] - E_{\text{tot}}[X].$$

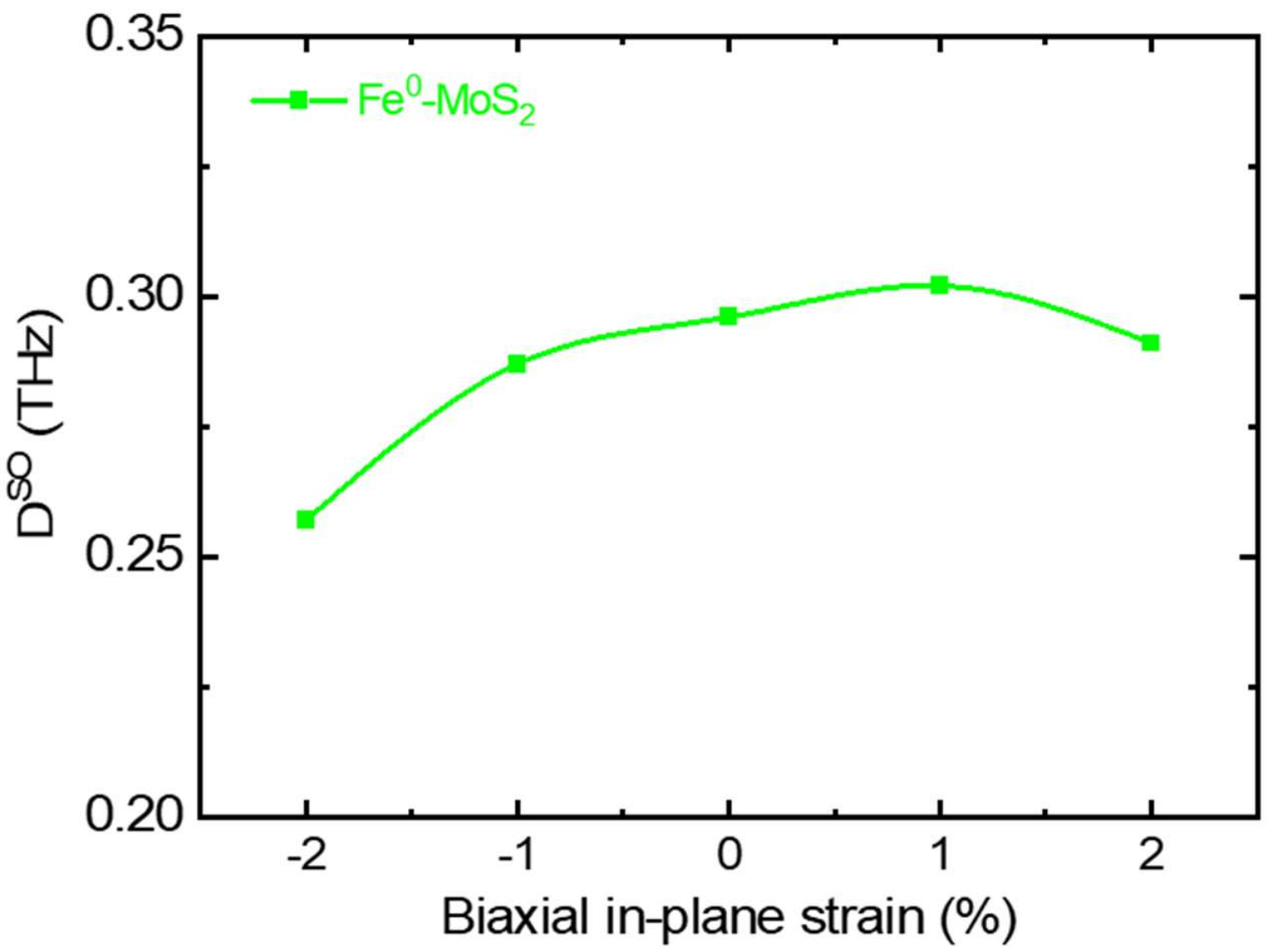

**Figure S2. ZFS value as a function of in-plane strain.** $D^{\mathrm{SO}}$ of $Fe^0$-$MoS_2$ vs biaxial in-plane strain from –2% to 2%.

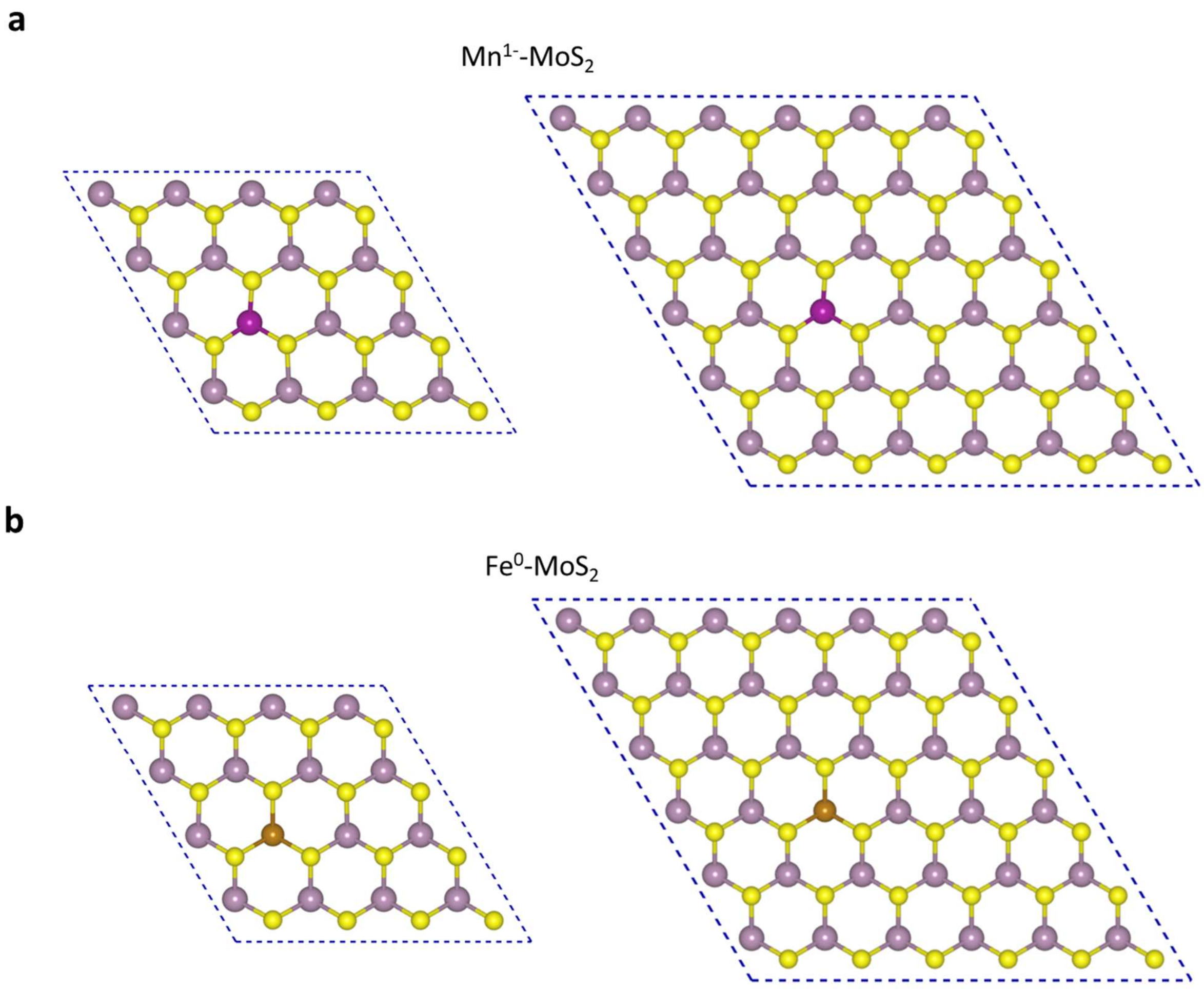

**Figure S3. Relaxed structures of (a) $Mn^{1-}$-$MoS_2$ and (b) $Fe^{0}$-$MoS_2$.** Left: 4×4 supercell; Right 6×6 supercell.

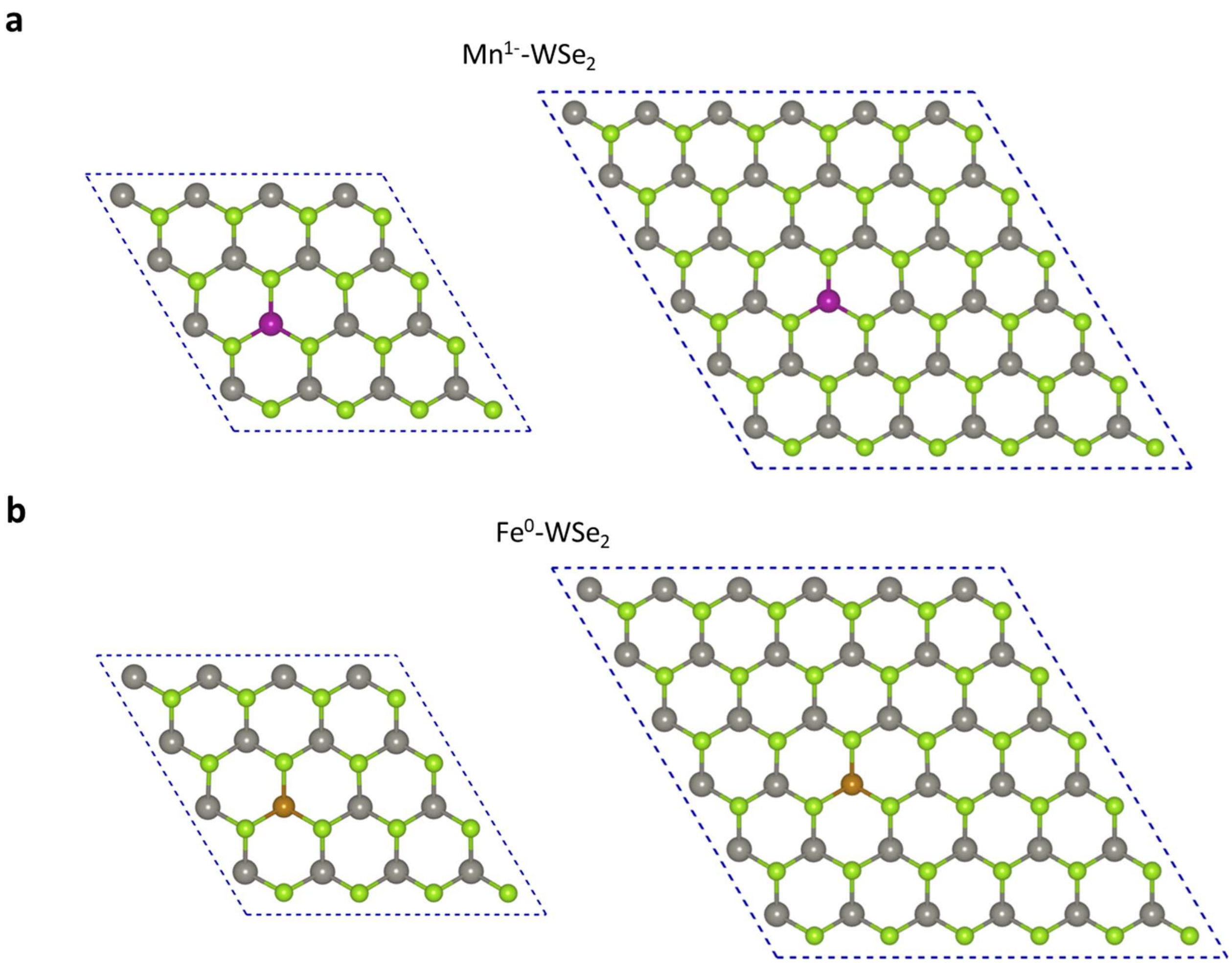

**Figure S4. Relaxed structures of (a) $Mn^{1-}$-$WSe_2$ and (b) $Fe^{0}$-$WSe_2$.** Left: 4×4 supercell; Right 6×6 supercell.

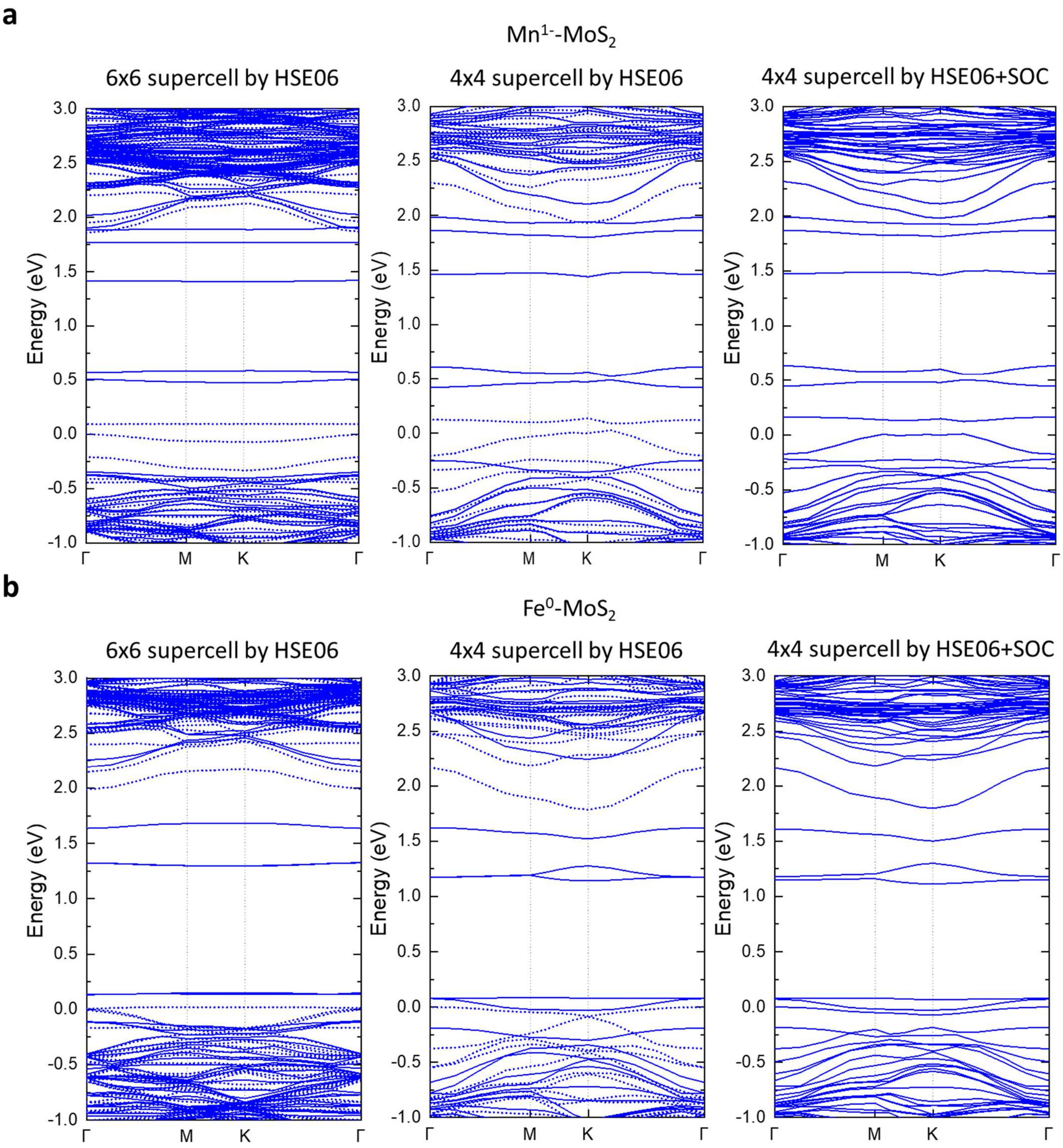

**Figure S5. Band structures for (a) $Mn^{1-}$-$MoS_2$ and (b) $Fe^{0}$-$MoS_2$.** Left:6×6 supercell without SOC; Middle: 4×4 supercell without SOC; Right: 4×4 supercell with SOC. The solid and dashed lines for band structures without SOC indicate the spin up and spin down states, respectively. The energies and spin states of the defect levels in the band gap are unchanged when compared to a 6×6 supercell or when compared to a band structure that includes the effects of SOC.

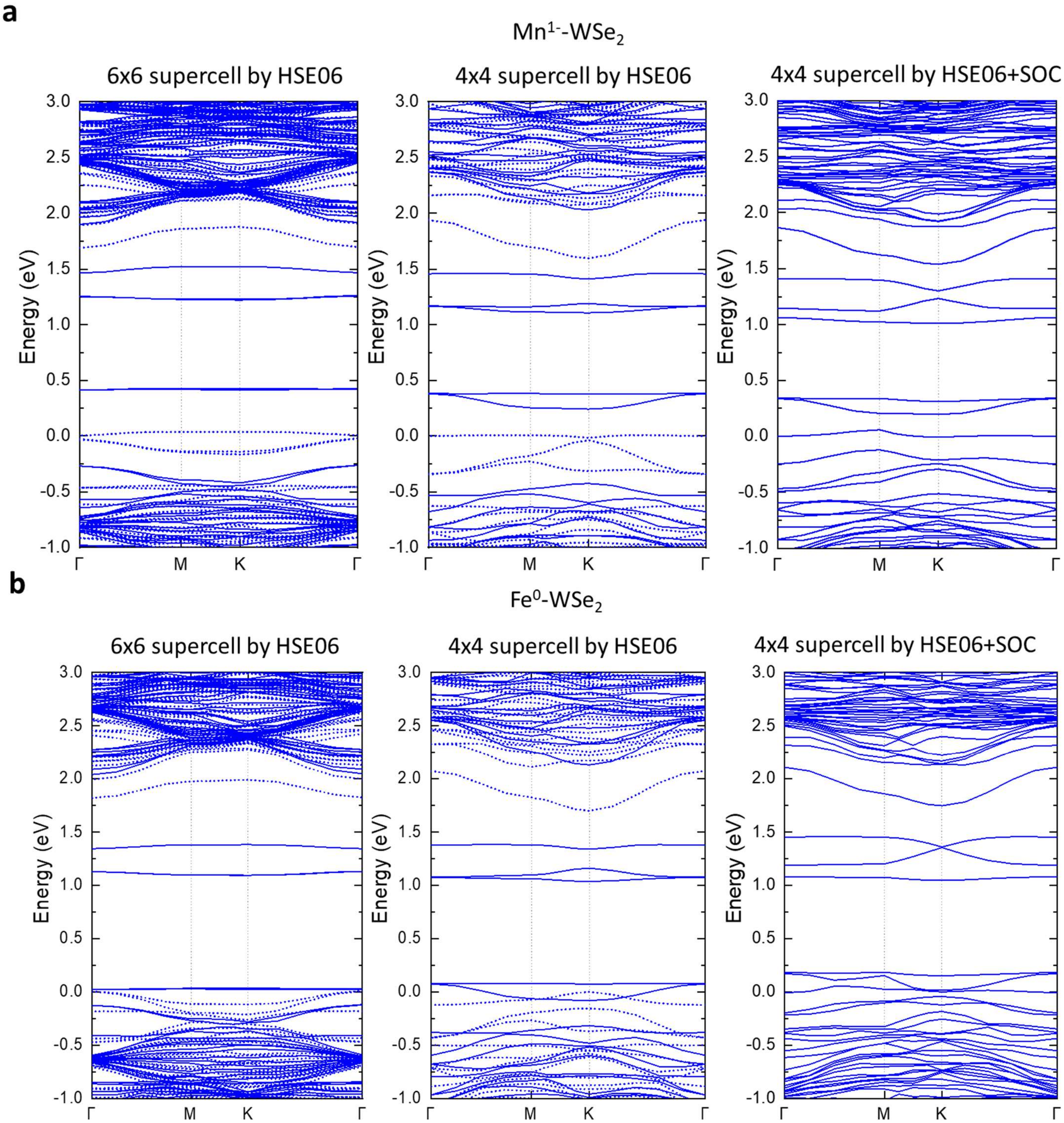

**Figure S6. Band structures for (a) $Mn^{1-}$-$WSe_2$ and (b) $Fe^{0}$-$WSe_2$.** Left:6×6 supercell without SOC; Middle: 4×4 supercell without SOC; Right: 4×4 supercell with SOC. The solid and dashed lines for band structures without SOC indicate the spin up and spin down states, respectively. The energies and spin states of the defect levels in the band gap are largely unchanged when compared to a 6×6 supercell or when compared to a band structure that includes the effects of SOC.

**Note S1: Symmetry-driven defect geometry optimization.**

Below are all the possible local defect symmetries with respect to dopant displacement vectors for initial disruptions.

| Symmetry | Displacement vector |
|---|---|
| $D_{3h}$ | 0 |
| $C_{3v}$ | $\vec{c}$ |
| $C_{2v}$ | (1) $-\vec{a}+\vec{b}$, (2) $\vec{a}+2\vec{b}$ |
| $C_s$ | (1) $-\vec{a}+\vec{b}+\vec{c}$, (2) $\vec{a}+2\vec{b}+\vec{c}$ |
| $C_h$ | $\vec{b}$ |
| $C_1$ | $\vec{b}+\vec{c}$ |

The corresponding schematic structures of ML TMDs with a substitutional TM dopant, top view (left) and side view (right). The shaded region indicates the possible displacement range of the dopant reduced by the symmetry.

For our initial screening calculations, we have found that, as a compromise between an exhaustive search and prudence in the use of computational resources, it is useful to relax using starting with the displacement vector for the lowest symmetry, $C_1$ for all the TM dopants in three charge states (+1, –1, and 0) with low and high spin multiplicities. In these calculations, PBE (+U for 3d elements) was used.

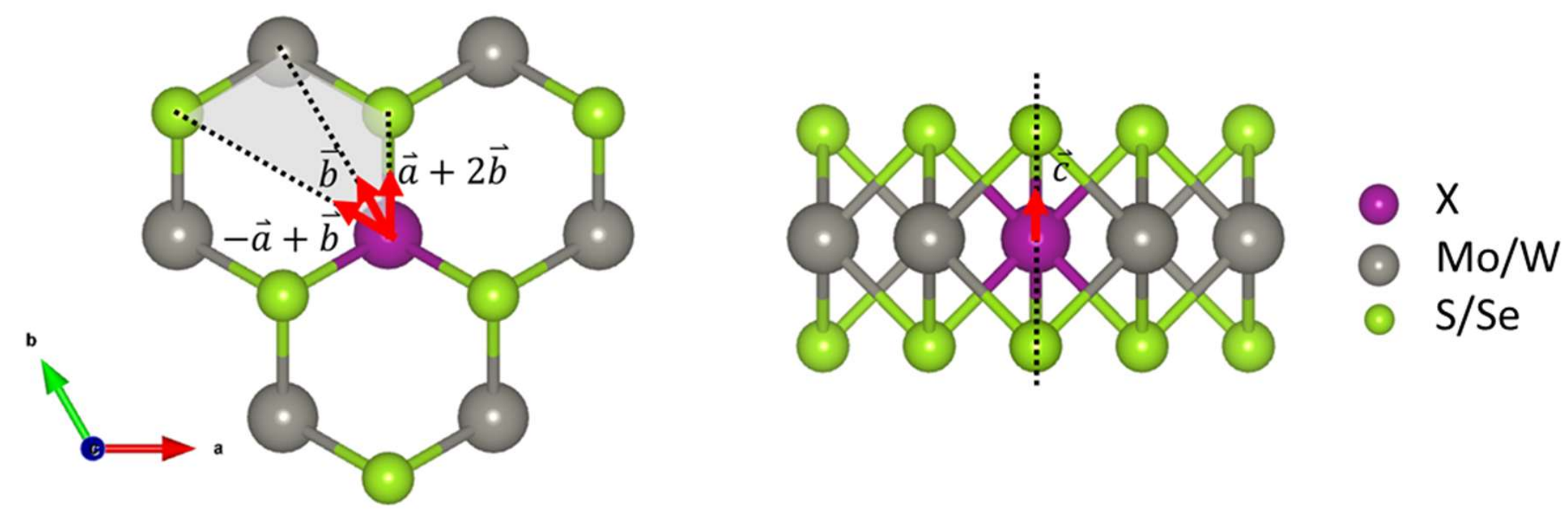

Selected systems of particular interest to us, e.g. those with stable triplet states, are then reassessed more carefully. In these cases, we explicitly consider initial geometries in the different symmetries in our geometry optimisation process. Initial and relaxed symmetries, spin magnetic moments, total energies of $Mn^{1-}$-$MoS_2$ and $Fe^{0}$-$MoS_2$ (PBE+U, U = 3 eV including SOC) are listed below. The final, most stable geometries, correspond to those that were obtained using our procedure above, relaxing from $C_1$.

**Relaxation for the defect structures of $Mn^{1-}$-$MoS_2$**

| Initial | Relaxed | Fractional coordinates | Spin magnetic moment ($\mu_B$) | Total energy (eV) |
|---|---|---|---|---|
| $D_{3h}$, $C_{3v}$ | $D_{3h}$ | 0.333, 0.417, 0.500 | 0.00, 0.00, 2.00 | -344.557 |
| $C_{2v}$ (1), $C_s$ (1)<br>$C_h$, $C_1$ | **$C_{2v}$ (1)** | 0.329, 0.422, 0.500 | 0.00, 0.00, 2.00 | **-344.608** |
| $C_{2v}$ (2), $C_s$ (2) | $C_{2v}$ (2) | 0.338, 0.426, 0.500 | 0.00, 0.00, 2.00 | -344.582 |

**Relaxation for the defect structures of $Fe^{0}$-$MoS_2$**

| Initial | Relaxed | Fractional coordinates | Spin magnetic moment (μB) | Total energy (eV) |
|---|---|---|---|---|
| $D_{3h}$, $C_{3v}$,<br>$C_{2v}$ (2), $C_s$ (2), $C_1$ | **$D_{3h}$** | 0.333, 0.417, 0.500 | -0.00, -0.00, 2.00 | **-342.404** |
| $C_{2v}$ (1), $C_s$ (1) | $C_{2v}$ (1) | 0.325, 0.425, 0.500 | -0.00, 0.00, 2.00 | -342.315 |
| $C_h$ | $C_h$ | 0.330, 0.428, 0.500 | 0.00, 0.00, 2.00 | -342.316 |

Then we further relax and verify the results with HSE06. The energy of spin-singlet state is found to much higher than that of the spin-triplet state. This result also agrees with the calculation where SOC is included, in which the spin magnetic moments are relaxed to the parallel spin with total magnetic moment of 2 $\mu_B$.

| **$Mn^{1-}$-$MoS_2$** | **Triplet Spin** | Triplet Spin | Singlet Spin |
|---|---|---|---|
| Relaxed symmetry | **$C_{2v}$** | $D_{3h}$ | $C_{2v}$ |
| Fractional coordinates | **0.328, 0.422, 0.500** | 0.333, 0.417, 0.500 | 0.326, 0.409, 0.500 |
| Total energy (eV) | **-417.426** | -417.246 | -416.736 |
| **$Fe^{0}$-$MoS_2$** | **Triplet Spin** | Triplet Spin | Singlet Spin |
| Relaxed symmetry | **$D_{3h}$** | $C_{2v}$ | $C_h$ |
| Fractional coordinates | **0.333, 0.417, 0.500** | 0.329, 0.421, 0.500 | 0.315, 0.408, 0.500 |
| Total energy (eV) | **-414.977** | -414.476 | -414.283 |

**Note S2: Details on the group theory analysis**

(i) Under group theory analysis for the $D_{3h}$ symmetry,[15] in the ground state, two electrons are in the $e''_x/e''_y$ degenerate orbitals. Therefore, the spatial part of the wavefunction transforms as $E''\otimes E'' = A'_1\oplus A'_2\oplus E'$ and the corresponding states are matched by the projection rule:[15]

$$A'_2: \frac{1}{\sqrt{2}}(e''_x e''_y - e''_y e''_x)$$

$$A'_1: \frac{1}{\sqrt{2}}(e''_x e''_x + e''_y e''_y)$$

$$E'_x: \frac{1}{\sqrt{2}}(e''_x e''_x - e''_y e''_y)$$

$$E'_y: \frac{1}{\sqrt{2}}(e''_x e''_y + e''_y e''_x)$$

The antisymmetric spatial part of spin-triplet state transforms as $A'_2$, while the rest of the three symmetric states form a spin-singlet state. In the first excited state (manifold), however, the electrons have IRs of $E''\otimes E' = A''_1\oplus A''_2\oplus E''$. As they do not occupy the same orbital, both symmetric (+) and antisymmetric (–) forms exist:[16]

$$A''_1(\pm): \frac{1}{2}(e'_x e''_y \pm e''_y e'_x - e'_y e''_x \mp e''_x e'_y)$$

$$A''_2(\pm): \frac{1}{2}(e'_x e''_x \pm e''_x e'_x + e'_y e''_y \pm e''_y e'_y)$$

$$E''_x(\pm): \frac{1}{2}(e'_x e''_x \pm e''_x e'_x - e'_y e''_y \mp e''_y e'_y)$$

$$E''_y(\pm): \frac{1}{2}(e'_x e''_y \pm e''_y e'_x + e'_y e''_x \pm e''_x e'_y).$$

The spin-triplet and spin-singlet states are assigned to every symmetric and antisymmetric function, respectively. Their degeneracy is lifted by the Coulomb repulsion energies.[16] Among the four spin-triplet sublevels, the ZPL transition can only occur between ground state $A'_2$ and excited state $A''_1(-)$ through $z$-polarized photons (see also the discussion in the main text).

The spin-part of the wavefunctions transforms as the direct product of the spin double group IR, $D_{1/2}\otimes D_{1/2}$, which is equivalent to the direct sum of representations corresponding to a spin-triplet and a spin-singlet state. In both cases, the states have integer values of spin angular momentum and thus transform as single group IRs (see Pg 353 of Ref. 15). The spin quantisation axis for this system is in the $z$-direction. The spin projections $\{S_z, S_x/S_y\}$ of the spin-triplet configuration transform as $\{R_z, R_x/R_y\}$. According to Ref. 15 (Pg 356), $m = +1$ can be identified with $S_+$, and $m = -1$ can be identified with $S_-$. Hence, the $|+\rangle$ state can be identified with $S_x$ and the $|-\rangle$ state with $S_y$. The spin projections on $\{S_z, S_x/S_y\}$ correspond to $\{A_2', E''\}$. The spin-singlet configuration corresponds to the fully symmetric IR $A_1'$.

For $Fe^0$-$MoS_2$, the total many-body wavefunction is configured by a direct product of the IRs of the spatial and spin part. Hence, the $S = 1$ ground state forms a manifold of ${}^3A_2'$ with IR of $E_{x,y}''$ states for $|\pm\rangle$ and $A_1'$ for a higher $|0\rangle$ state. The manifold of lowest excited-state ${}^3A_1''$ has the higher $|0\rangle$ state with IR of $A_2''$, and the lower $|\pm\rangle$ states with IR of $E_{x,y}'$. Similar analysis can be applied for $Mn^{-1}$-$WSe_2$.

(ii) For $Fe^0$-$MoS_2$, the singlet state has a slight Jahn-Teller (JT) distortion towards $C_{2v}$ symmetry, which is a subgroup of $D_{3h}$ ($\sigma_h \rightarrow \sigma_{xy}$). The selection rules governing coupling between the singlet state and the ground or excited triplet states can thus be inferred from the correlation between the IRs of $D_{3h}$ and its $C_{2v}$ subgroup.

| **$D_{3h}$** | $A_1$' | $A_2$' | E' | A"$_1$ | A"$_2$ | E" |
|---|---|---|---|---|---|---|
| **$C_{2v}$** | $A_1$ | $B_1$ | $A_1 + B_2$ | $A_2$ | $B_1$ | $A_2 + B_1$ |

(iii) For $Mn^{1-}$-$MoS_2$ with JT-distorted $C_{2v}$ symmetry (character table listed below), the two highest occupied state $b_2$ and $1a_1$ configure the spatial part of ground state through $b_2 \otimes a_1 = b_2$. Likewise, the first excited state (manifold) transforms as $B_2$ as well. For the ground state, the spin quantisation axis is in the $z$-direction, and the spin part of the wavefunctions for $\{|0\rangle, |+\rangle, |-\rangle\}$ can be assigned the IRs $\{B_2, B_1, A_2\}$ transforming as $\{R_z, R_x, R_y\}$ in the $C_{2v}$ point group. Together with the spatial part of the wavefunction, the total many-body wavefunction thus transforms as $\{A_1, A_2, B_1\}$. For the excited state, the spin quantisation axis is in the $x$-direction, and in this case, the spin part of the wavefunctions for $\{|0\rangle, |+\rangle, |-\rangle\}$ can be assigned the IRs $\{B_1, A_2, B_2\}$ transforming as $\{R_x, R_y, R_z\}$ in the $C_{2v}$ point group. The total many-body wavefunction the spin-triplet state (both ground and first-excited state) forms a manifold of ${}^3B_2$ with IR of $\{A_1, A_2, B_1\}$.

| **$C_{2v}$** | **E** | **$C_2$(y)** | **$\sigma_v$(zy)** | **$\sigma_v$(xy)** | **Linear** | **Quadratic** |
|---|---|---|---|---|---|---|
| $A_1$ | 1 | 1 | 1 | 1 | y | $x^2$;$y^2$;$z^2$ |
| $A_2$ | 1 | 1 | -1 | -1 | $R_y$ | zx |
| $B_1$ | 1 | -1 | 1 | -1 | z; $R_x$ | zy |
| $B_2$ | 1 | -1 | -1 | 1 | x; $R_z$ | xy |